\documentclass[preprint,12pt]{elsarticle}
\DeclareUnicodeCharacter{2212}{-}
\usepackage[a4paper, lmargin=0.1666\paperwidth, rmargin=0.1666\paperwidth, tmargin=0.1111\paperheight, bmargin=0.1111\paperheight]{geometry} 

\usepackage{amssymb}
\usepackage[fleqn]{amsmath}
\usepackage{derivative}
\usepackage[hyphens]{url}
\usepackage[hidelinks]{hyperref}
\usepackage{booktabs}
\newcommand{\del}[1]{}
\usepackage{xcolor, soul}
\sethlcolor{yellow}
\usepackage{mdframed}
\usepackage{multicol}
\usepackage{multirow}
\usepackage{float} 
\usepackage{geometry}
\usepackage{caption}
\usepackage{bm}
\usepackage{dutchcal,verbatim}
\usepackage[american]{circuitikz}
\ctikzset{bipoles/resistor/width=.7}
\ctikzset{bipoles/resistor/height=.36}
\ctikzset{
sources/fill=pink, 
csources/fill = cyan,}
\newcommand{\n}[1]{_{\mathrm{#1}}}

\newcommand{\Cw}{C\n{w}}
\newcommand{\Cc}{C\n{c}}
\newcommand{\Rce}{R\n{ce}}
\newcommand{\Rwa}{R\n{wa}}
\newcommand{\Rcw}{R\n{cw}}
\newcommand{\Ta}{T\n{a}}
\newcommand{\Te}{T\n{e}}
\newcommand{\Tc}{T\n{c}}
\newcommand{\Tw}{T\n{w}}

\journal{}
\usepackage{xcolor, soul}
\sethlcolor{white}
\begin{document}
\captionsetup[figure]{labelfont={},name={Fig.},labelsep=period}
\captionsetup[table]{labelfont={},name={Table},labelsep=period}
\begin{frontmatter}



\title{A step towards digital operations - A novel grey-box approach for modelling the heat dynamics of Ultra-low temperature freezing chambers}

\author[inst1]{Tao Huang\corref{cor1}}
\cortext[cor1]{Corresponding author}\ead{taohu@dtu.dk}
\affiliation[inst1]{organization={Technical University of Denmark, Department of Applied Mathematics and Computer Science},
            addressline={Asmussens Allé, Building 303B}, 
            city={Kgs. Lyngby},
            postcode={8000}, 
            country={Denmark}}

\author[inst1]{Peder Bacher}
\author[inst1]{Jan Kloppenborg Møller}
\author[inst2]{Francesco D’Ettorre}
\author[inst2]{Wiebke Brix Markussen}
\affiliation[inst2]{organization={Danish Technological Institute},
            addressline={Gregersensvej 1}, 
            city={ Taastrup},
            postcode={2630}, 
            country={Denmark}}

\begin{abstract}
Ultra-low temperature (ULT) freezers store perishable bio-contents and have high energy consumption, which highlight a demand for reliable methods for intelligent surveillance and smart energy management. This study introduces a novel grey-box modelling approach based on stochastic differential equations to describe the heat dynamics of the ULT freezing chambers. The proposed modelling approach only requires temperature data measured by the embedded sensors and uses data from the regular operation periods for model identification. \hl{The model encompasses three states: chamber temperature, envelope temperature, and local evaporator temperature. Special attention is given to the local evaporator temperature state, which is modelled as a time-variant system, to characterize the time delay and dynamic variations in cooling intensity.}\del{The model has three states, of which a time-variant model with nonlinear input for the local evaporator temperature state is specifically established to adapt to the variation of the cooling intensity at the position of the embedded chamber control probe.} Two ULT freezers with different operational patterns are modelled. The unknown model parameters are estimated using the maximum likelihood method. The results demonstrate that the models can accurately predict the chamber temperature measured by the control probe \hl{(RMSE $<$ 0.19 $^{\circ}$C)} and are promising to be applied for forecasting future states. In addition, the model for local evaporator temperature can effectively adapt to different operational patterns and provide insight into the local \del{evaporation}\hl{cooling supply} status. The proposed approach greatly promotes the practical feasibility of grey-box modelling of the heat dynamics for ULT freezers and \del{is a step forward in future digital operations}\hl{can serve several potential digital applications. A major limitation of the modelling approach is the low identifiability, which can potentially be addressed by inferring model parameters based on relative parameter changes. }
\end{abstract}


\end{frontmatter}
\begin{figure}[H]
	\centering
\begin{mdframed}
	\begin{multicols}{2}
		\textbf{Nomenclature}
		\begin{description}
             \item[\textnormal{\emph{Variables and parameters}}]
                \item[\textnormal{$C^\text{c}$}] Heat capacity of the chamber state
                \item[\textnormal{$C^\text{e}$}] A parameter determining the inertia of the evaporator
                \item[\textnormal{$C^\text{w}$}] Heat capacity of the envelope state
                \item[\textnormal{$m_t$}] Compressor state signal (0/1)
                \item[\textnormal{$M_t$}] \del{Modified}\hl{Transformed} compressor state signal (-1/1)    
                \item[\del{\textnormal{$m_t^\text{ac}$}}\hl{\textnormal{$M_t^\text{ac}$}}] Accumulated \hl{transformed} compressor state signal 
                \item[\textnormal{$R^\text{ce}$}] Thermal resistance between the chamber and local evaporator
                \item[\textnormal{$R^\text{cw}$}] Thermal resistance between the chamber and envelope
                \item[\textnormal{$R^\text{wa}$}] Thermal resistance between the envelope and ambient environment
                \item[\textnormal{$T_t^\text{a}$}] Ambient air temperature ($^{\circ}$C)
			\item[\textnormal{$T_t^\text{c}$}] Chamber temperature measured by RTD ($^{\circ}$C)
                \item[\textnormal{$T_t^\text{e}$}] Local evaporator temperature ($^{\circ}$C)
   			\item[\textnormal{$T_t^\text{e,in}$}] Evaporator inlet temperature ($^{\circ}$C)
 
      		\item[\textnormal{$T_t^\text{e,out}$}] Evaporator outlet temperature ($^{\circ}$C)
             \item[]        \item[]
                \item[\textnormal{$T_t^\text{w}$}] Chamber envelope temperature ($^{\circ}$C)
                \item[\hl{\textnormal{$S_t(\cdot)$}}] Sigmoid function (from 0 to 1)
                \item[\textnormal{$\omega_t$}] Standard Wiener process
       
                \item[\textnormal{$\bm{X}_k$}] The stochastic state variable of the system at time $t_k$
                \item[\textnormal{$\bm{Y}_k$}] The stochastic observation variable of the system at time $t_k$
                \item[\textnormal{$\mathcal{L}(\cdot)$}] Likelihood function
                \item[\textnormal{$\mathcal{L}_\text{P}(\cdot)$}] Profile likelihood function
                \item[\textnormal{$\mathcal{l}(\cdot)$}] Log-likelihood function
                \item[\textnormal{$\mathcal{l}_\text{P}(\cdot)$}] Profile log-likelihood function
                                  \item[]
                \item[\textnormal{\emph{Abbreviations}}]
                \item[\textnormal{ACF}] Auto-correlation Function
                \item[\textnormal{FDD}] Fault Detection and Diagnostics
                \item[\textnormal{F\#1}] Freezer No. 1
                \item[\textnormal{F\#2}] Freezer No. 2
                \item[\textnormal{MPC}] Model Predictive Control
                \item[\textnormal{M$_{F\#1}$}] Model for Freezer No. 1
                \item[\textnormal{M$_{F\#2}$}] Model for Freezer No. 2
                \item[\textnormal{RTD}] Resistance Thermal Detector
                \item[\textnormal{ULT}] Ultra-low Temperature 
                \item[\textnormal{2CRS}] Two-stage Cascading Refrigeration System

		\end{description}
	\end{multicols}
\end{mdframed}
\end{figure}

\newpage
\section{Introduction}
\del{Ultra-low temperature (ULT) freezers play an essential role in pharmaceutical businesses and research organizations. It is also gained increasing deployments in society during the COVID-19 pandemic for storing vaccines \cite{pfizer}. The contents, such as vaccines, tissues, blood samples, and organs. are most often of a high monetary value and a fast-degradable biological nature. Continuous surveillance of ULT freezing chamber temperature thus is essential to secure content quality and enable early fault detection and diagnosis (FDD).}

\hl{Ultra-low temperature (ULT) freezers play a vital role in pharmaceutical businesses and research organizations by preserving valuable and perishable items such as tissues, blood samples, and organs. Their significance was further underscored during the COVID-19 pandemic when ULT freezers played a critical role in storing vaccines \mbox{\cite{pfizer}}. To ensure the quality of stored contents, it is essential to have continuous monitoring and early fault detection and diagnosis (FDD) of the thermal condition in freezing chambers.} 

\hl{Compared to standard refrigerators, ULT freezers operate at -40 $^{\circ}$C to -90 $^{\circ}$}C. To reach low temperatures,\del{ ULT freezers are commonly equipped with a complex cascading refrigeration system  \cite{review}To reach low temperatures. } \hl{an ULT freezer may consume up to 20 kWh/day \mbox{\cite{Uedb}}. This is estimated to be three times more than the daily power consumption of an average Danish household \mbox{\cite{el}}. Consequently, ULT freezers are among the most energy-intensive pieces of equipment in hospitals, bio-banks, laboratory buildings, etc. \mbox{\cite{doi:10.1504/WRSTSD.2013.050786, KU}}. Given the growing emphasis on the transition to low-carbon cities, it is also imperative to improve their energy efficiency and unlock energy flexibility \mbox{\cite{OCONNELL2014686, DETTORRE2022112605, FDULT}}}. \del{A}\hl{A promising solution is to replace the simple set-point control with advanced model predictive control (MPC) \mbox{\cite{mpc, YANG2021117112}}.}\del{At the same time, the primary function of securing the quality of the contents must be guaranteed. This necessitates continuous monitoring of the status of the ULT freezing chambers in order to provide basic operation information, fault detection and diagnosis (FDD), and corrective recommendations for the end users.}

\hl{However, an essential prerequisite to achieving continuous monitoring, FDD, and MPC in ULT freezers is the availability of a} reliable dynamic model\del{is essential for achieving these targets}. Previous studies have attempted to model the heat dynamics of refrigerator chambers using different approaches. They can be categorized as white-box modelling \cite{MASTRULLO201438, ZSEMBINSZKI2017188}, black-box modelling \cite{LI2022114, 10.1145/2768510.2768536, https://doi.org/10.1002/er.1218}, and grey-box modelling. The \hl{deterministic} white-box models are established based on detailed physical descriptions of the systems without including stochastic model parts. They are physically plausible but usually demanding to build and are not useful for statistical estimations of parameters. Black-box models are based purely on data-driven approaches. They are superior in computational load at the expense of physical interpretability. \hl{Stochastic} grey-box modelling is an intermediate approach between white\hl{-} and black-box modelling. Grey-box models are formulated based on physical considerations but explicitly include stochastic descriptions of unexplained variation in data. The models preserve critical physical meanings and leverage the advantage of pure data-driven approaches. The parameters can be estimated and tracked over time, which makes it possible to detect systematic changes linked directly to the operational status. Thus, grey-box models have better generalization properties than other types of models \cite{AFRAM2014507} and have proven to be more \del{suitable}\hl{reliable} for practical implementations \cite{ZONG20171476, DECONINCK2016290, Thilker2021, ROUCHIER2018181}. 

Several grey-box models for refrigerator chambers can be found in the literature. O’Neill et al. \cite{ONEILL2014819} modelled the heat dynamics of cold food storage rooms based on ordinary differential equations (ODEs). Leerbeck et al. \cite{9698514, LEERBECK2023100211} used stochastic differential equations (SDEs) to formulate the models of the heat dynamics of supermarket refrigerator cabinets. Sossan et al. \cite{SOSSAN20161} and Costanzo et al. \cite{6604197} modelled the heat dynamics of household refrigerator chambers and \del{tested}\hl{evaluated} the potential to use the model for simple MPC purposes. These models effectively capture the heat dynamics of the cold chambers. However, they rely on high-quality data from controlled experiments, which are intentionally designed to excite the systems for model identification. In addition, these models require various physical quantities as inputs apart from the temperatures, such as refrigerant flow rate, pressures at critical locations, and\del{power rate.} \hl{power consumption.} These attributes demand \del{extra}sophisticated sensors and are difficult to measure non-invasively from the refrigeration loops. All these issues challenge the practical feasibility of these models. Even so, the literature on dynamic modelling for ULT freezers is more scarce than for standard refrigerators. \hl{Existing} studies \del{about ULT freezers}primarily focus on the pull-down performance \cite{LIU2023105537, tan}, theoretical thermodynamics analyses \cite{BHATTACHARYYA20091077, KILICARSLAN20102947, SUN201619, SUN20191170}, and the effects of working fluids \cite{DOPAZO2011257, SANZKOCK201441, LLOPIS2015133, ZHU2021114380}. Due to \del{different}\hl{varying} settings and operational conditions, it is imperative to explore valid approaches for modelling the heat dynamics of ULT freezers. 

Owing to business-critical contents, the \hl{modern} ULT freezers are often outfitted with multiple temperature sensors \hl{strategically placed in the chamber and at the critical points along the refrigeration loops} to facilitate maintenance and trigger warm alarms. \del{In most cases, a precise temperature probe is installed in the chamber to control the compressors, while several sensors are placed at the critical points along the refrigeration loops. }This setup generates a vast amount of real-time temperature data from \hl{the} regular operation. Considering the practical feasibility, a modelling approach, which exclusively relies on the costless data from the embedded sensors in ULT freezers, is more promising to be generalized and implemented in practice\del{accepted by the end-users}. 
\subsection{Challenges of using the measurements from the embedded sensors for grey-box modelling}\label{sec:dif}
Lumped resistance and capacitance (RC) models are a popular and effective approach to establishing grey-box models. When considering the freezing chamber as a lumped system, the temperature measured in the core region is considered representative of the global thermal condition in the chamber. In typical ULT freezers without assisting fans, the heat transfer within the chamber is dominated by natural convection, resulting in an inhomogeneous \del{vertical}\hl{spatial} temperature distribution \cite{LIU2023105537, tan}. Therefore, the temperature probe used for control purposes is usually attached to the surface of the warmest part to ensure that the entire chamber is sufficiently cold, typically in the lower part of the chamber. This is mainly because the refrigerant typically enters the evaporation coil from the upper part of the freezer and exits as superheated gas from the lower part \hl{and the refrigerant used in the evaporator part of the cycle is usually a mixture temperature having a temperature glide during evaporation}. Therefore, the lower refrigeration coil has \del{the lowest heat transfer rate, leading to }a higher temperature, \hl{leading to a warmer} lower chamber surface. Thus, the temperature measured by the control probe is not necessarily the mean chamber temperature but tends to be a local temperature. It has been reported in \cite{tan} that this temperature near the lower chamber surface exhibits a distinct temporal evolution pattern compared to the core region, and its dynamic response is not aligned with the states (ON/OFF) of the compressors. This may imply underlying nonlinear dynamics during heat transfer in local regions and poses the first challenge for modelling. In addition, the natural convection causes the local surface temperature to be determined by the cooling intensity from the proximate evaporation coil. However, the cooling capacity can vary along the long coil, from the refrigerant inlet to the outlet. Consequently, the cooling input to a specific local chamber region is not constant and can vary under different practical scenarios. Thus, the model must have the capability to adapt to such local variation. This presents the second challenge when directly using the signals from the embedded sensors for grey-box modelling. Besides, the model must have acceptable complexity and a high generalisation potential.

\subsection{Objectives}
To this end, this study presents a grey-box modelling approach for describing the heat dynamics of the ULT freezing chamber, which has not been considered previously. The proposed grey-box modelling approach only relies on costless data from the embedded sensors. \del{and does not require demanding measurements.}Specifically, a novel time-variant model for the local evaporator temperature is developed. The major novelty is the nonlinear part, where a time delay is modelled in continuous time. The modelling results are validated against the data from the regular operation period and residual analysis. Furthermore, we assess the model identifiability through profile likelihood and discuss a possible strategy for inferring the critical parameters to compensate for the low physical interpretability. The results from this study intend to contribute to a more practical grey-box modelling approach and thus be of use in future digital operation and dynamical monitoring of ULT freezers.

This paper is structured as follows: Section \ref{sec:data} provides information about the modelled ULT freezers and the data sources. Section \ref{sec:methods} describes the modelling \del{processes, which includes}\hl{procedures, including} detailed information about the model structure and system identification technique. The \del{issue regarding parameter inference}\hl{method for assessing the model identifiability} is also outlined in \hl{Section \mbox{\ref{sec:methods}}}\del{this section}. \hl{The results, including}\del{ Section \ref{sec:results} presents the results regarding model performance,} parameter estimates, \hl{unconditional predictions, }residual analysis\hl{, and long-term model performance, along with the corresponding discussions,} are presented in Section \ref{sec:results}. \hl{This section also discusses the potential strategy for parameter interpretation, possible model applications, and}  associated limitations\del{ are also discussed}. Conclusions are followed in Section \ref{sec:conclusions}.
\section{Descriptions of the ULT freezers and data}\label{sec:data}
\begin{figure}[H]
\centering
\includegraphics[width=0.4\linewidth]{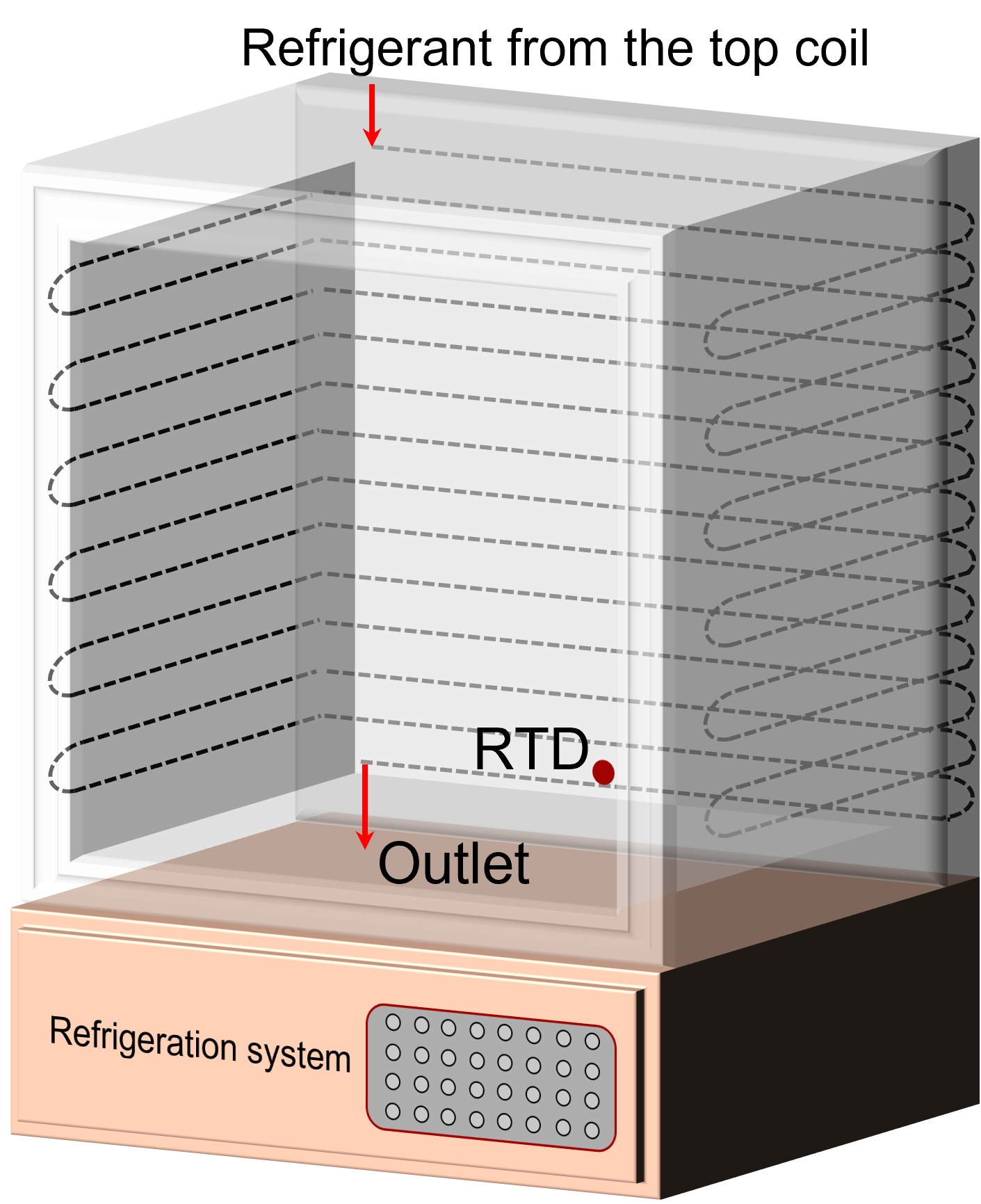}
\caption{Schematic of the layout of the evaporation coil and RTD. This figure is used as an illustration, so there may be a deviation from the actual conditions. The top coils are not drawn.}
\label{Freezer}
\end{figure}

Two ULT freezers in regular daily use are selected for modelling in this study. Both freezers are commercial products from the same manufacturer (Thermo Scientific). Freezer No. 2 (F\#2) is the predecessor \del{product}of Freezer No. 1 (F\#1). The selected freezers are equipped with a 2-stage cascading refrigeration system \hl{(2CRS)} \cite{review} and have an identical interior chamber dimension of 1300 $\times$ 686 $\times$ 1019 mm, with four shelves. The states of the compressors (ON/OFF) are regulated based on a thermostatic control according to the temperature level measured by a Resistance Temperature Detector (RTD) mounted on the backside of the lowest shelf, see Fig. \ref{Freezer}. The evaporation coil uniformly wounds the freezing chamber, except for the doors and bottom surfaces. The cold refrigerant flows into the evaporation coils from top to bottom of the chamber. The internal heat transfer is predominantly governed by natural convection and radiation\del{ due to the absence of assisting fans in the chamber} \hl{since there are no assisting fans}. Despite these similarities, the two freezers differ in the refrigerant mixtures and the types of condensers. Because of the different operational patterns, we modelled both ULT freezers to investigate if the proposed modelling strategy can be generalised to freezers with different settings. The freezers are operated by the same end-user\del{s}. The defrosting of the freezer is conducted manually on a fixed schedule.
\subsection{The data}
\begin{figure}[H]
\centering
\includegraphics[width=1\linewidth]{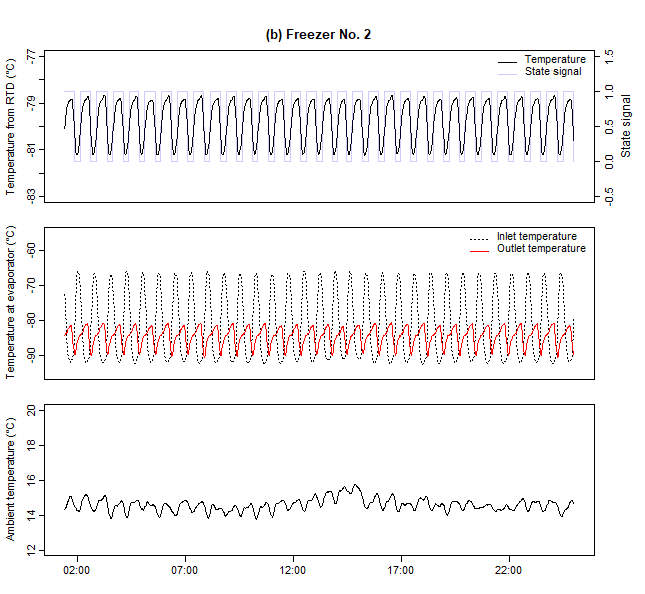}
\caption{Temperature profiles of the selected freezers during regular operation periods.}
\label{F1}
\end{figure}

The selected ULT freezers were situated in a room with a controlled ambient temperature. The data for modelling was recorded during regular operation periods. The thermostatic set-point of the freezers was at -80 $^{\circ}$C. A total of eight temperatures were measured. Besides RTD, the rest of the sensors were standard thermal couples and measured the inter-stage heat exchanger temperature, 1$^\text{st}$- and 2$^\text{nd}$-stage suction temperatures, and 2$^\text{nd}$-stage sump temperature. Considering the model complexity, we did not model the detailed refrigeration systems and selected four temperatures for modelling, as listed in Table \ref{T1}. The chamber temperature measured by the RTD is used to represent the chamber thermal condition and is hereafter referred to as RTD temperature. Although the RTD does not measure the mean chamber temperature, it is more reasonable to be used for modelling because 1) it represents the warmest, thus the worst condition, 2) the freezer is controlled based on this level, and 3) placing the temperature sensor in the core region of the chamber \hl{could reduce space efficiency and thus may not always be practical.} The ambient temperature $T^\text{a}$ was measured at the condenser air inlet. In addition to the temperatures, the compressor running state $m$ was registered and expressed as a binary signal\del{, i.e.,} (OFF = 0, ON = 1). The compressor is turned on when the RTD temperature is above the set-point\del{and turned off vice versa}. This is the built-in control. Events such as door opening/closing were also logged. The temperature sensors and data logging systems are already embedded into the freezers by the manufacturer. All measurements were sampled at 1 min resolution. During the data recording period, the freezers were not emptied.\del{The loading contents of the freezers are unknown.}
\begin{table}[H]
\caption{Variables used for grey-box modelling.}
\renewcommand{\baselinestretch}{1}
\small
\centering
\begin{tabular}{l l l}
\toprule 
\textbf{Variable}& \textbf{Description}& \textbf{Unit}\\
\midrule
$T^\text{c}$ & Chamber temperature measured by the RTD & \multirow{4}{*}{$^{\circ}$C} \\
$T^\text{e,in}$&Evaporator inlet temperature&\\
$T^\text{e,out}$&Evaporator outlet temperature&\\
$T^\text{a}$&Ambient temperature&\\
$m$ & State signal& 0 / 1\\
\bottomrule
\end{tabular}
\label{T1}
\end{table}
Fig. \ref{F1} illustrates the time series of the temperature signals measured from the two freezers during undisturbed periods for 24 h \hl{for model identification}. Undisturbed periods mean that the freezers were operating without any events occurring. The primary difference between the temperature pattern of F\#1 and F\#2 is the temporal evolution of the RTD temperature during the ON state. For F\#1, the RTD temperature responds fast to the compressor state signal. The span of the pull-down duration is similar to that of the ON state. In contrast, the response of the RTD temperature of F\#2 to the state signal is \del{cryptic}complex. When the compressors are ON, the RTD temperature rises for some time before suddenly dropping sharply. The pull-down duration of the RTD temperature is much shorter than the span of the ON state. In addition, there exists a time lag between the shifts in the compressor state signal and the response of the RTD temperature. The presented temperature profiles for F\#1 and F\#2 are found to be typical for the same generation of units in relation to the compressor state signals. \del{All t}These phenomena imply that there may be \del{some }nonlinear dynamical behaviours on the cooling transport from the refrigeration loops to the local freezing chamber. Thus, directly using the compressor state signal in the model would be insufficient to express the variation in the cooling capacity. \hl{Transformations of the state signals must be performed to tackle this issue and} should be discretely considered during the modelling processes. Moreover, the profiles of the evaporator inlet and outlet temperatures also differ markedly between the two freezers. 
\del{The temperature difference between the evaporator inlet and outlet in F\#1 becomes stable shortly after the shifts from OFF to ON states, indicating that steady-state operation can quickly reach. However, the evolution of the evaporator inlet and outlet temperatures in F\#2 suggests that a stable temperature level is barely reached within the ON state. This may consequently lead to a continuous change in the intensity of the boiling effect.} The \del{overall }discrepancies between F\#1 and F\#2 could be attributed to the different settings in the refrigeration loops and actual operating conditions. Such discrepancies reveal that the operational patterns can vary significantly among ULT freezers. This highlights the imperative of developing a highly adaptive modelling approach for ULT freezers.
\begin{figure}[H]
\centering
\includegraphics[width=1\linewidth]{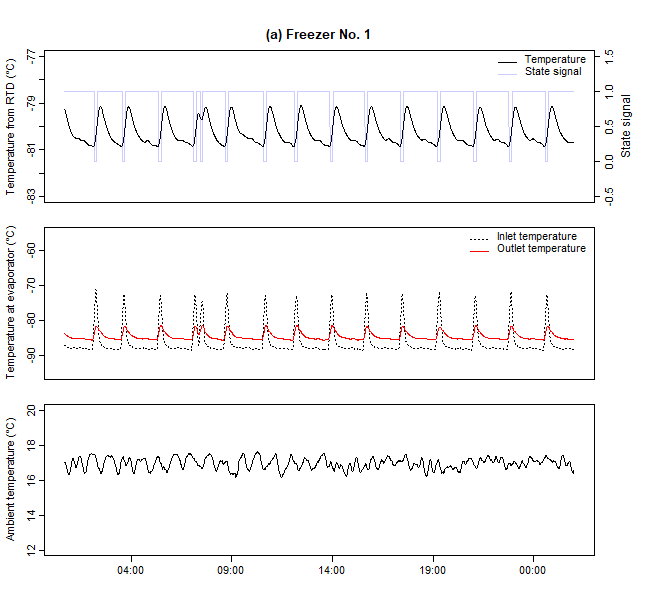}
\end{figure}
\section{Grey-box model development}\label{sec:methods}
This section describes the grey-box modelling procedures. In previous models for cold \hl{storage} spaces \cite{LEERBECK2023100211, SOSSAN20161}, the exact cooling capacity or power consumption is commonly used as an input. However, flow/pressure or energy meters are not expected to be largely deployed in commercial ULT freezers because they are costly and may result in refrigerant leakage. In most prevailing cases, temperature signals are the only available measurements. Thus, the proposed approach aims to establish a model based solely on temperature measurements.
\subsection{Stochastic differential equation}
The proposed grey-box model is formulated using SDEs in continuous time state-space representation. The basic model structure is
\setlength{\mathindent}{0cm}
\begin{equation}\label{eq:1}
d\bm{X}_t = \underbrace{f\bm{(X}_t,\bm{U}_t,t, \bm{\theta})dt}_\text{\clap{Drift term ~}} + \underbrace{\bm{\sigma}_t(\bm{\theta})d\bm{\omega}_t}_\text{\clap{~ Diffusion term}},
\end{equation}
where, $t\in \mathbb{R}$ is time, $\bm{X}_t \in \mathbb{R}^\text{N}$ the state vector and $\bm{U}_t \in \mathbb{R}^\text{m}$ is the input vector. $\bm{\theta} \in \mathbb{R}^\text{p}$ is the parameter vector. The diffusion term differentiates the stochastic calculus from ODEs. It represents the system error and is modelled as a random walk mimicking the $n$-dimensional Standard Weiner processes. $\bm{\sigma}_t$ is the associated standard deviation. Compared to ODEs, grey-box modelling based on SDEs is more comprehensive because the diffusion terms compensate for the modelling approximations, unrecognised and unmodeled inputs by describing the stochastic nature of random errors in observed data \cite{MADSEN199567}. 
\subsection{Basic model for heat dynamics of ULT freezing chambers}
The heat transfer between the ULT freezing chamber and the refrigeration system is mainly driven by the temperature gradient. The decay processes of the RTD temperature shown in Fig. \ref{F1} exhibit \hl{a pattern with a rapid initial decrease followed by a slow decrease. This suggests that the dynamics of the chamber temperature can be better captured by a second-order model, considering the presence of at least two time constants.} Thus, in addition to the state of the RTD temperature, we model the chamber wall temperature as a hidden state, such that it serves as a low-pass filter between the ambient environment and the freezing chamber. The fundamental heat dynamic model of the freezing chamber is thus formulated as 
\begin{align}\label{eq:3}
dT_t^\text{c} &=\frac{1}{C^\text{\hl{c}}}\left(\frac{T_t^\text{w}-T_t^\text{c}}{R^\text{\hl{cw}}}+\frac{T_t^\text{e}-T_t^\text{c}}{R^\text{\hl{ce}}}\right)dt+\sigma^{\text{c}}d\omega_t^\text{c}, 
\end{align}
\begin{align}\label{eq:4}
dT_t^\text{w} &=\frac{1}{C^\text{\hl{w}}}\left(\frac{T_t^\text{a}-T_t^\text{w}}{R^\text{\hl{wa}}}+\frac{T_t^\text{c}-T_t^\text{w}}{R^\text{\hl{cw}}}\right)dt+\sigma^{\text{w}}d\omega_t^\text{w},
\end{align}
where, $T^\text{c}$ is the RTD temperature state. $T^\text{w}$ is the chamber wall temperature state. $C^\text{c}$ and $C^\text{w}$ are the heat capacities of the states. $R^\text{cw}$, $R^\text{ce}$, and $R^\text{wa}$ are the thermal resistances between different lumped parts. All these parameters are unknown and need to be estimated. $T^\text{e}$ denotes the evaporator temperature and can be regarded as the cooling input in the system. Like the RTD temperature, $T^\text{e}$ does not necessarily correspond to the mean evaporator temperature, but rather the local evaporator temperature near the RTD. Measuring this temperature can be difficult because the evaporation coil is \hl{sealed behind the chamber casing}. We thus model $T^\text{e}$ as another hidden state by leveraging the available temperature measurements at the evaporator inlet and outlet. 
\subsection{The time-variant dynamic model for local evaporator temperature}
The local evaporator temperature $T^\text{e}$ is modelled as an individual hidden state written as a differential equation on itself
\begin{equation}\label{eq:6}
dT_t^\text{e} =\frac{1}{C^\text{\hl{e}}}\left((\underbrace{aT^\text{e,out}_t + bT^\text{e,in}_t}_{\text{\hl{hypothetical} evaporator temperature}}-T_t^\text{e})\cdot \underbrace{S_t(M_t^\text{ac})}_{\text{Sigmoid function}}\right)dt+\sigma_\text{e}d\omega_t^\text{e}.
\end{equation}
The drift part of the model is elaborated in the following contents. The misalignment observed between the response of the RTD temperature and the compressor state signal shown in Fig. \ref{F1} suggests a time-delayed nonlinearity in the heat transfer. Describing this feature requires addressing two issues. First, there is \del{an exponential dynamic response} \hl{a time delay} between the shift of the state signal and temperature drops. Such a response can be well interpreted by the thermal inertia of the evaporator, i.e., the evaporator needs to be cooled down before the heat in the chamber can be absorbed. This indicates that a buffer factor needs to be introduced to govern the inertia. We denote this factor as an unknown parameter $C^\text{e}$. 

\begin{figure}[H]
\centering
\includegraphics[width=1\linewidth]{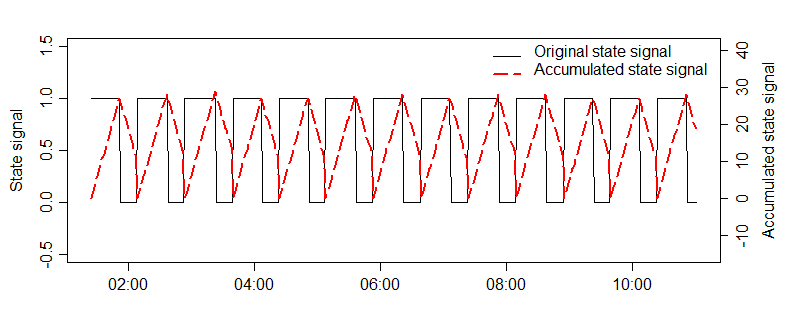}
\caption{An example of accumulated transformed state signal compared to the original state signal.}
\label{mac}
\end{figure}

Second, the span of the ON state is different from that of the pull-down duration, which is particularly evident for F\#2  (see Fig. \ref{F1} (b)), where the temperature drops very suddenly, and the pull-down period is much shorter than the ON state time. This indicates that the cooling capacity takes some time to reach the nominal condition, and such a process appears not linear. In previous studies \cite{LEERBECK2023100211, SOSSAN20161}, the response of the chamber temperature is well correlated with the state signal in both starting time and duration. The discrepancy might be mainly caused by the \hl{reduced} cooling capacity in the local evaporation coil as introduced in Subsection \ref{sec:dif}. It might also be partially due to the\del{sophisticated} dual-system settings from the 2CRS. To account for the local nonlinear effect, we first conceived a \hl{hypothetical} evaporator temperature, approximated as a linear combination of the evaporator inlet and outlet temperatures, i.e., $aT^\text{e,out}_t$ + $bT^\text{e,in}_t$. The inlet temperature represents the evaporator temperature at nominal cooling capacity, while the outlet temperature is the super-heated refrigerant temperature. Depending on the scales of $a$ and $b$, the resulting \hl{hypothetical} evaporator temperature tends to be an intermediate temperature biased either towards the evaporator outlet or inlet, which appears reasonable. Technically, the sum of $a$ and $b$ should be close to 1. We intend to use this \hl{hypothetical} evaporator temperature to represent an imaginary local cold source\del{ temperature}. The \hl{hypothetical} temperature equals the actual local evaporator temperature if and only if the linear relationship holds. This condition is not always met because both latent and sensible heat exchanges happen in the evaporator. Therefore, a novel approach with a sigmoid function is applied to model the nonlinearity of the heat transfer on the local evaporator. \hl{The Sigmoid function writes as}
\begin{equation}\label{eq:5}
S_t(M_t^\text{ac}) = \frac{1}{1+\exp(-\alpha(M_t^\text{ac}-\beta))}.
\end{equation}
A sigmoid function has a characteristic "S"-shaped curve and is parameterized by $\alpha$ and $\beta$, which determine the slope and the offset, respectively. Its output is bounded in the range of 0 to 1. Eq. (\ref{eq:5}) has one variable $M_t^\text{ac}$ = $\Sigma_{t_{m_{t = 0}}} ^{t_m{_{t=1}}} M_t $, the accumulated transformed state signal $M_t$ within a duty cycle. $M_t$ is 1 or -1 according to the binary compressor state signals $m_t$, as
\begin{equation}\label{eq:M}
    M_t = \Bigg\{
        \begin{array}{ll}
        1, & \text{if } m_t = 1,\\
        -1, & \text{if } m_t = 0.
        \end{array}
\end{equation}
The accumulated variable $M_t^\text{ac}$ is reset to zero at the moment when the state signal shifts from 0 to 1, as exemplified in Fig. \ref{mac}. 

The parameters $\alpha$ and $\beta$ need to be estimated along with other unknown parameters. By using appropriate values of $\alpha$ and $\beta$, the sigmoid function modifies the trajectory of the compressor state signal, resulting in a \del{desired}nonlinear smooth transition. Furthermore, we assumed that the transition between ON and OFF follows a similar variation profile. This means the same sigmoid function is applied regardless of the state of the compressors. This allows both a smooth decay and a fast increase of the modelled local evaporator temperature, as is expected in practice. Overall, the sigmoid function regulates the changing rate of the local evaporator temperature $T_t^\text{e}$ towards the \hl{hypothetical}\del{local} evaporator temperature\del{ $aT^\text{e,out}_t$ + $bT^\text{e,in}_t$}, such that the local \hl{temporal} variation in the cooling intensity and nonlinear heat transfer can be described. 

\hl{Eventually, the full model of the freezing chamber is formulated by Eqs. (\mbox{\ref{eq:f1}-\ref{eq:f3}})}. \del{By introducing the sigmoid function}\hl{Since none of the model parameters depends on the state variables}, the model describes a $3^\text{rd}$-order linear time-variant system.
\begin{align}\label{eq:f1}
dT_t^\text{c} &=\frac{1}{C^\text{c}}\left(\frac{T_t^\text{w}-T_t^\text{c}}{R^\text{cw}}+\frac{T_t^\text{e}-T_t^\text{c}}{R^\text{ce}}\right)dt+\sigma^{\text{c}}d\omega_t^\text{c}, 
\end{align}
\begin{align}\label{eq:f2}
dT_t^\text{w} &=\frac{1}{C^\text{w}}\left(\frac{T_t^\text{a}-T_t^\text{w}}{R^\text{wa}}+\frac{T_t^\text{c}-T_t^\text{w}}{R^\text{cw}}\right)dt+\sigma^{\text{w}}d\omega_t^\text{w},
\end{align}
\begin{align}\label{eq:f3}
dT_t^\text{e} =\frac{1}{C^\text{e}}\left((aT^\text{e,out}_t + bT^\text{e,in}_t-T_t^\text{e})\cdot S_t(M_t^\text{ac})\right)dt+\sigma^\text{e}d\omega_t^\text{e}.
\end{align}

\hl{The Sigmoid function $S_t(M_t^\text{ac})$ represents the time-variant parameter in the system. Considering it as a whole, the model can be written as the following matrix form}
\begin{multline}\label{eq:ma}
\begin{bmatrix}
dT_t^\text{c} \\ dT_t^\text{w} \\ dT_t^\text{e} 
\end{bmatrix} 
=
\begin{bmatrix}
-\frac{1}{C^\text{c}R^\text{cw}}-\frac{1}{C^\text{c}R^\text{ce}} & \frac{1}{C^\text{c}R^\text{cw}} & \frac{1}{C^\text{c}R^\text{ce}} \\
\frac{1}{C^\text{w}R^\text{cw}} & -\frac{1}{C^\text{w}R^\text{wa}}-\frac{1}{C^\text{w}R^\text{cw}} & 0\\
0 & 0 & -\frac{S_t(M_t^\text{ac})}{C^\text{e}}
\end{bmatrix}
\begin{bmatrix}
T_t^\text{c} \\ T_t^\text{w} \\ T_t^\text{e} 
\end{bmatrix}
dt+ \\
\begin{bmatrix}
0&  0 & 0\\
\frac{1}{C^\text{w}R^\text{wa}} & 0& 0\\
0& a\frac{S_t(M_t^\text{ac})}{C^\text{e}} & b\frac{S_t(M_t^\text{ac})}{C^\text{e}}
\end{bmatrix}
\begin{bmatrix}
T_t^\text{a} \\ T^\text{e,out}_t \\T^\text{e,in}_t 
\end{bmatrix}
+
\begin{bmatrix}
 \sigma^\text{c} & 0 & 0 \\ 0 & \sigma^\text{w} & 0 \\0 & 0 & \sigma^\text{e}
\end{bmatrix}
\begin{bmatrix}
d\omega_t^\text{c}\\ d\omega_t^\text{w} \\ d\omega_t^\text{e}
\end{bmatrix},
\end{multline}
\hl{where, the system state vector consists of $[T_t^\text{c}, T_t^\text{w}, T_t^\text{e}]^\text{T}$. The model input vector, the second term on the right-hand side of Eq. (\mbox{\ref{eq:ma}}), consists of $[T_t^\text{a}, T^\text{e,out}_t, T^\text{e,in}_t ]^\text{T}$. In addition, $M_t^\text{ac}$ is required to calculate the Sigmoid function. Therefore, the model has a total of four inputs \{$T_t^\text{a}, T^\text{e,out}_t, T^\text{e,in}_t, m_t$\} for performing predictions. All these four variables are continuously registered by the embedded sensors and logging systems in the ULT freezers.}

\begin{figure}[H]
  \centering
  \newcommand{\scaleTikz}{1}
  \tikzset{every node/.style={scale=\scaleTikz}}
  \begin{circuitikz}[xscale=\scaleTikz, yscale=\scaleTikz]
    \def\x{3}    
    \def\y{3}    
    \def\ylab{4} 
    \coordinate (Evaporator) at (0,0);
    \coordinate (Chamber) at ($(Evaporator)+(\x,0)$);
    \coordinate (Wall) at ($(Chamber)+(\x,0)$);
    \coordinate (Ambient) at ($(Wall)+(\x,0)$);
    \draw
    ($(2*\x,0)$) node[ground] {}
    (Evaporator) -- (Ambient) 
    ($(Evaporator)+(0,\y)$) -- ($(Evaporator)+(0,\y)$); 
    \draw to [csV, l=$\Te$] ($(Evaporator)+(0,\y)$) to (Evaporator);
    \draw ($(Evaporator)+(0,\y)$) to [R, l=$\Rce$] ($(Chamber)+(0,\y)$);
    \def\xoffset{0.1*\x}
    \draw ($(Chamber)+(\xoffset,0)$) to[C=$\Cc$, -*] ($(Chamber)+(\xoffset,\y)$)
    {[anchor=south] ($(Chamber)+(0,\y)$) node {$\Tc$}};
    \draw ($(Chamber)+(0,\y)$) to [R, l=$\Rcw$] ($(Wall)+(0,\y)$);
    \def\xoffset{0.1*\x}
    \draw ($(Wall)+(\xoffset,0)$) to[C=$\Cw$,  -*] ($(Wall)+(\xoffset,\y)$)
     {[anchor=south] ($(Wall)+(0,\y)$) node {$\Tw$}};
    \draw ($(Wall)+(0,\y)$) to [R, l=$\Rwa$] ($(Ambient)+(0,\y)$);
    \draw [V, l=$\Ta$] ($(Ambient)+(0,\y)$) to (Ambient);
   \draw
      [dashed] ($(Evaporator)+(0.7*\x,-0.7)$) -- ($(Evaporator)+(0.7*\x,\ylab)$)
        [dashed] ($(Chamber)+(0.7*\x,-0.7)$) -- ($(Chamber)+(0.7*\x,\ylab)$)  
     [dashed] ($(Wall)+(0.8*\x,-0.7)$) -- ($(Wall)+(0.8*\x,\ylab)$); 
    \draw
        {[anchor=south] ($(Evaporator)+(0,\ylab)$) node{Evaporator}}
        {[anchor=south west] ($(Chamber)+(-0.5,\ylab)$) node{Chamber}}
       {[anchor=south west] ($(Wall)+(0,\ylab)$) node{Wall}}
       {[anchor=south west] ($(Ambient)+(0,\ylab)$) node{Ambient}};
  \end{circuitikz}
  \caption{Equivalent RC-network of the 3-state model. The modelled \hl{local} evaporator temperature \hl{$T^\text{e}$} acts as a controlled source.}
  \label{RC}
\end{figure}
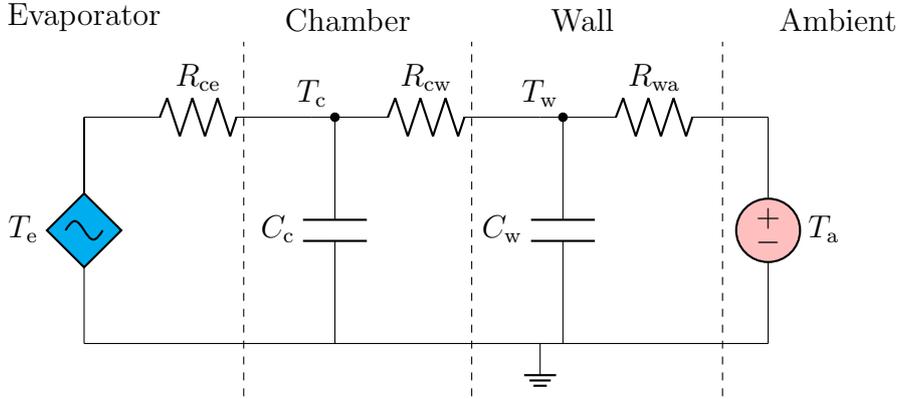

We only observe $T^c$, so the observation equation is expressed as
\begin{equation}\label{eq:7}
y_t =T_t^\text{c}+e_t,
\end{equation}
where, $e_t \sim N(0, \nu)$ is the independent and identically distributed random error from the measurements. Fig. \mbox{\ref{RC}} shows the simplified equivalent RC circuit of the \hl{3-state} system.
\subsection{Model identification and parameter estimation}
The model parameters are estimated using maximum likelihood estimation. Given the observation sequence $\bm{\mathcal{Y}}_N$ = $\{{y}_k\}_{k=0}^N$ and a vector $\bm{\theta}$ contains all model parameters and variances, the likelihood function is the joint probability density function (\del{pdf}\hl{PDF}) of the observations assuming that all parameters are known, expressed as
\begin{equation}\label{eq:8}
\mathcal{L}(\bm{\theta};\bm{\mathcal{Y}}_\text{N} ) = p(y_\text{0}|\bm{\theta})(\prod_\text{t=1}^Np(y_k|\bm{\mathcal{Y}}_{k-1}),\bm{\theta}),
\end{equation}
where, $p(y_\text{0}|\bm{\theta})$ is the likelihood at initial state $k = 0$. $p(y_k|\bm{\mathcal{Y}}_{k-1}),\bm{\theta})$ is the conditional \del{pdf}\hl{PDF} of observing $y_{k}$ given the previous observations and the model parameters. Because the increments of $\bm{\omega}$ and \del{$\bm{\nu}$}\hl{${\nu}$} are normally distributed white noises, the conditional probability density is in the model assumed to follow the Gaussian distribution, which can be characterized by the mean and variance. Therefore, the likelihood function and be calculated using the Gaussian density function, given the one-step prediction error $\epsilon_k$
\begin{equation}\label{eq:9}
\epsilon_k = y_k - \hat{y}_{k|{k-1}},
\end{equation}
where, $\hat{y}_{k|{k-1}}$ is the expectation of observing $y_k$ given one-step ahead observation $y_{k-1}$ and model parameters, i.e., $\hat{y}_{k|{k-1}}$ =  E$[y_k|y_{k-1}, \bm{\theta}]$. The associated variance is thus $R_{k|{k-1}}$ =  Var$[y_k|y_{k-1}, \bm{\theta}]$. The log-likelihood function is commonly used to facilitate numerical calculations, as
\begin{equation}\label{eq:10}
\mathcal{l}(\bm{\theta};\bm{y}_N) = \log(p(y_\text{0}|\bm{\theta}))-((\frac{1}{2}\Sigma_{k=0}^N\frac{\epsilon_k^2}{R_{k|{k-1}}}+\log(\sqrt{2\pi \cdot R_{k|{k-1}} })).
\end{equation}
Eventually, the model parameters are determined by minimizing the negative log-likelihood function
\begin{equation}\label{eq:minl}
\hat{\theta} = \operatorname*{arg\,min}_{\bm{\theta}} -\mathcal{l}(\bm{\theta};\bm{y}_N).
\end{equation}
The log-likelihood function is evaluated by calculating the one-step prediction errors $\epsilon_k$ and the associated variance $R_k$. The Kalman filtering approach can be used for this purpose to estimate the states. For systems with a nonlinear input, we employ the continuous-discrete extended Kalman filter (EKF) by linearizing the estimation using partial derivatives of the system and observation functions. Please refer to \cite{kF1, KF2, KRISTENSEN2004225} for details about EKF. The data processing and scientific computing are conducted in the programming language R.
\subsection{Model identifiability}\label{sec:pi}
The proposed grey-box model is purely informed by temperature signals. It has no dimension constraint for the thermal parameters, and the model parameters are estimated based on non-intrusive data. Consequently, the model identifiability is undermined, and some estimated parameters may lack absolute physical meanings. To assess the identifiability of the model, \hl{we calculate the correlations between estimated parameters. The calculation procedures can be found in \mbox{\ref{apped:a}}. In addition,} we \hl{also} use profile likelihood ($\mathcal{L}_\text{P}$). The profile likelihood  is defined as the maximum likelihood estimated at fixed values of the parameters of interest
 \cite{LM}
\begin{equation}\label{eq:pl}
\mathcal{L}_\text{P}(\bm{\lambda};\bm{y}_N) = \operatorname*{sup}_{\bm{\zeta}} \mathcal{L}(\bm{\theta}; \bm{y}_N),
\end{equation}
where, $\bm{\theta}=(\bm{\lambda},\bm{\zeta})$ and $\bm{\lambda}$ ($p$-dimensional) denotes the parameters of interest. 

The profile likelihood for a practically identifiable model would show a concentrated bell-shaped distribution and an asymptotic confidence interval for all parameters. For a model lacking identifiability, the profile likelihood distribution is anticipated to be flat, meaning different sets of parameters can result in a similar likelihood. Consequently, the parameters may not be able to be associated with any physical proprieties in an absolute way. The profile likelihood-based 95\% confidence interval is defined as
\begin{equation}\label{eq:plci}
\{\bm{\lambda};\frac{\mathcal{L}_\text{P}(\bm{\lambda};\bm{y}_N)}{\mathcal{L}(\bm{\theta};\bm{y}_N)} > \exp(-\frac{1}{2}\chi^2_{0.95}(p))\} \Rightarrow  \{\bm{\lambda};\mathcal{l}_\text{P}(\bm{\lambda};\bm{y}_N) > \mathcal{l}(\bm{\theta};\bm{y}_N)-(\frac{1}{2}\chi^2_{0.95}(p))\},
\end{equation}
where, $\mathcal{l}_\text{P}$ is the profile log-likelihood.

\del{\subsection{Strategy for parameter interpretation}
In practical applications, the exact values of the parameters may not be of interest for appliances such as freezers, as their state suffers a continuous change over the entire lifetime. Activities such as maintenance or loading can easily alter the heat dynamics of the chamber. Consequently, some parameters can vary over time. For instance, the heat capacity of the chamber can change due to changes in the content mass after loading and unloading. The frosting on the inner surface of the chamber would increase the thermal resistance between the interior and the evaporator. As a result, the model needs to be updated to recapture the heat dynamics of the freezing chamber when the previous dynamical equilibrium is changed. Otherwise, the model prediction will drift from the measurements. 

Because the proposed model is based on the heat transfer theory, the relative changes in the parameters enable the possibility of physically interpreting the parameters relative to their past states, notwithstanding the lack of physical meanings. This can lead to a semi-interpretable grey-box model. From the event logs, we found that the model predictions could drift significantly after some door-opening events, which can be naturally related to loading/unloading behaviours. We thus selected a sufficiently long period from F\#2 that included door-opening events in order to illustrate how the critical parameter can possibly be interpreted to provide a relative implication for the operation status. However, the increased thermal resistance due to frosting could not be identified from the available data and was unfortunately not logged. This might be because the defrosting of the selected freezers was always conducted in a timely manner.} 
\section{Results and discussions}\label{sec:results}
\subsection{Parameter estimates}\label{sec:pe}
Table \ref{par} presents the estimated parameters for the models for F\#1 (M$_{F\#1}$) and for F\#2 (M$_{F\#2}$). All parameters are statistically significant, except for $C^\text{e}$ in M$_{F\#2}$ ($p \approx 0.1$). $C^\text{e}$ governs the hysteresis between the modelled evaporator and RTD temperatures. Thus, it is believed to be necessary for the model. The weak significance might be due to the limited information contained in the data regarding the dynamics of the local evaporator temperature and the poor quadratic approximation of the log-likelihood, leading to errors in the Wald Confidence Interval. It could be found that the scales of some of the parameters differ markedly between the two freezers. For instance, $R^\text{wa}$ for M$_{F\#1}$ is much higher than that for M$_{F\#2}$, while $R^\text{cw}$ for M$_{F\#1}$ is much smaller than that for M$_{F\#2}$. Similar issues can be found in $C^\text{c}$ and $C^\text{w}$. In a physical consideration, a plausible explanation for this can be that the heat resistance $R^\text{wa}$ for M$_{F\#1}$ is the total resistance of the freezing chamber and the interior contents. In contrast, the heat resistance of the interior contents is included in $R^\text{cw}$ for M$_{F\#2}$. The same explanation also applies to the discrepancy in $C$ values. In a statistical consideration, this implies that some parameters are not independent, and the models lack identifiability. Thus, the parameters may not be physically interpretable and should not be directly compared across models\del{ direct comparisons of parameters from models for different freezers may not be valid}. This issue will be further discussed in Subsection \ref{sec:ident}.
\begin{table}[H]
\caption{Estimated parameters for Model No. 1 and No. 2.}
\renewcommand{\baselinestretch}{1}
\small
\centering
\begin{tabular}{l l l l l}
\toprule 
\textbf{Parameter}& \textbf{M$_{F\#1}$}& \textbf{95\% CI}& \textbf{M$_{F\#2}$}& \textbf{95\% CI}\\
\midrule
$a$ & 4.78e-5&[4.0e-5, 5.6e-5]& 0.83 &[0.83, 0.84]\\
$b$&0.98&[0.98, 0.99]&0.12&[0.12, 0.13]\\
$C^\text{c}$&1.54&[0.46, 1.62]&15.58&[8.37, 22.78]\\
$C^\text{w}$&11.53&[8.09, 14.97]&0.35&[0.26, 0.44]\\
$C^\text{e}$&0.11&[9.8e-2, 0.11]&2.9e-3&[-5.9e-4, 6.4e-3]\\
$R^\text{wa}$ & 13.38&[12.05, 14.70]&0.30&[0.21, 0.39]\\
$R^\text{ce}$ & 0.55&[0.55, 0.56]& 1.5e-2&[0.01, 0.02]\\
$R^\text{cw}$ &0.20&[0.19, 0.21] & 9.30&[5.37, 13.23]\\
$\alpha$ & 0.37&[0.30, 0.43]& 0.21&[0.17, 0.26]\\
$\beta$ & 4.96&[4.45, 5.46]& 25.77&[16.88, 34.67]\\
\bottomrule
\end{tabular}
\label{par}
\end{table} 
 
It is worth discussing the parameters $a$ and $b$ because they somewhat indicate the \hl{mixture refrigerant properties, thereby the} level of the cooling intensity. In both models, $a+b \approx 1$, suggesting that $a$ and $b$ are self-bounded and negatively correlated. $a$ is so small that to be insignificant for M$_{F\#1}$, suggesting that the local evaporator temperature is very close to the \del{evaporator }inlet temperature. This indicates that the evaporation effect is still intensive \del{as it passes through the coil }near the RTD. In contrast, $a = 0.83$ in M$_{F\#2}$, indicating that the \hl{local} evaporator temperature around the RTD is closer to the outlet temperature of the superheated gaseous refrigerant. Therefore, the cooling intensity \del{at the evaporation coil }around the RTD is significantly reduced. Hence, \del{the heat transfer between the evaporator and the lower freezing chamber occurs mainly in a sensible form. As a result, }the lower chamber \del{ does not receive the nominal cooling capacity}\hl{receives reduced cooling capacity}. This phenomenon partially explains the mismatch between the compressor state signal and RTD temperature evolution in $F\#2$. 

In addition, \hl{the large discrepancy between $a$ and $b$ indicates that the local evaporator temperature does not reflect the average evaporation temperature, supporting that} the RTD temperature may not represent the mean chamber temperature. This can be a common feature for ULT freezers without assisting fans\del{, which can homogenise the temperature distribution and dynamic responses in the chamber}, in which the RTD temperature can often be higher than the mean chamber temperature due to reduced cooling capacity\del{ depends more on the cold evaporation coil at proximity. When the RTD is placed near the \hl{evaporator} outlet\del{ of the evaporator}, the RTD temperature is often higher than the mean chamber temperature due to reduced cooling capacity. Such a location for the control probe can be well understood from the manufacturers' perspective, that is, to ensure the temperature at the warmest place }\del{is lower than }\del{remains below the setpoint}.\del{However, this biased placement } This, on the one hand, leads to modelling challenges as mentioned in Subsection \ref{sec:dif}, and precludes the possibility of using global information\del{, such as nominal cooling capacity and compressor state signal,} directly as model inputs. One the other hand, it implicates that modelling global heat dynamics may risk an underestimation of the \del{thermal condition }\hl{temperatures} in local regions of the freezing chamber.\del{ This which should be avoided when monitoring the thermal condition in ULT freezers. } \hl{Thus, it is crucial to avoid using such models for monitoring the thermal condition in ULT freezers.}

Furthermore, the level of cooling inputs near the RTD is expected to vary under different practical scenarios\hl{, which can be attributed to variations in the mass quantities and spatial distribution of the loaded contents.}\del{This is due to the fact that the boundary between the dry region and the 2-phase region in the evaporation coil can move within a duty cycle. Moreover, The heat capacity and the spatial distribution of the loaded contents may also influence the variation of cooling intensity from the evaporator inlet to the outlet. This phenomenon} \hl{Describing such variations using partial differential equations with detailed boundary conditions would be possible but significantly increase the model complexity}\del{ may be explicitly described using partial differential equations given detailed boundary conditions. However, it would significantly increase the model complexity. In this light, t}. The proposed approach overcomes these issues by adjusting $a$ and $b$ values to adapt to different practical scenarios. This approach ameliorates the representation of the local evaporator state without over-complicating the model.
\subsection{Unconditional predictions}\label{sec:sim}
\begin{figure}[H]
\centering
\includegraphics[width=1\linewidth]{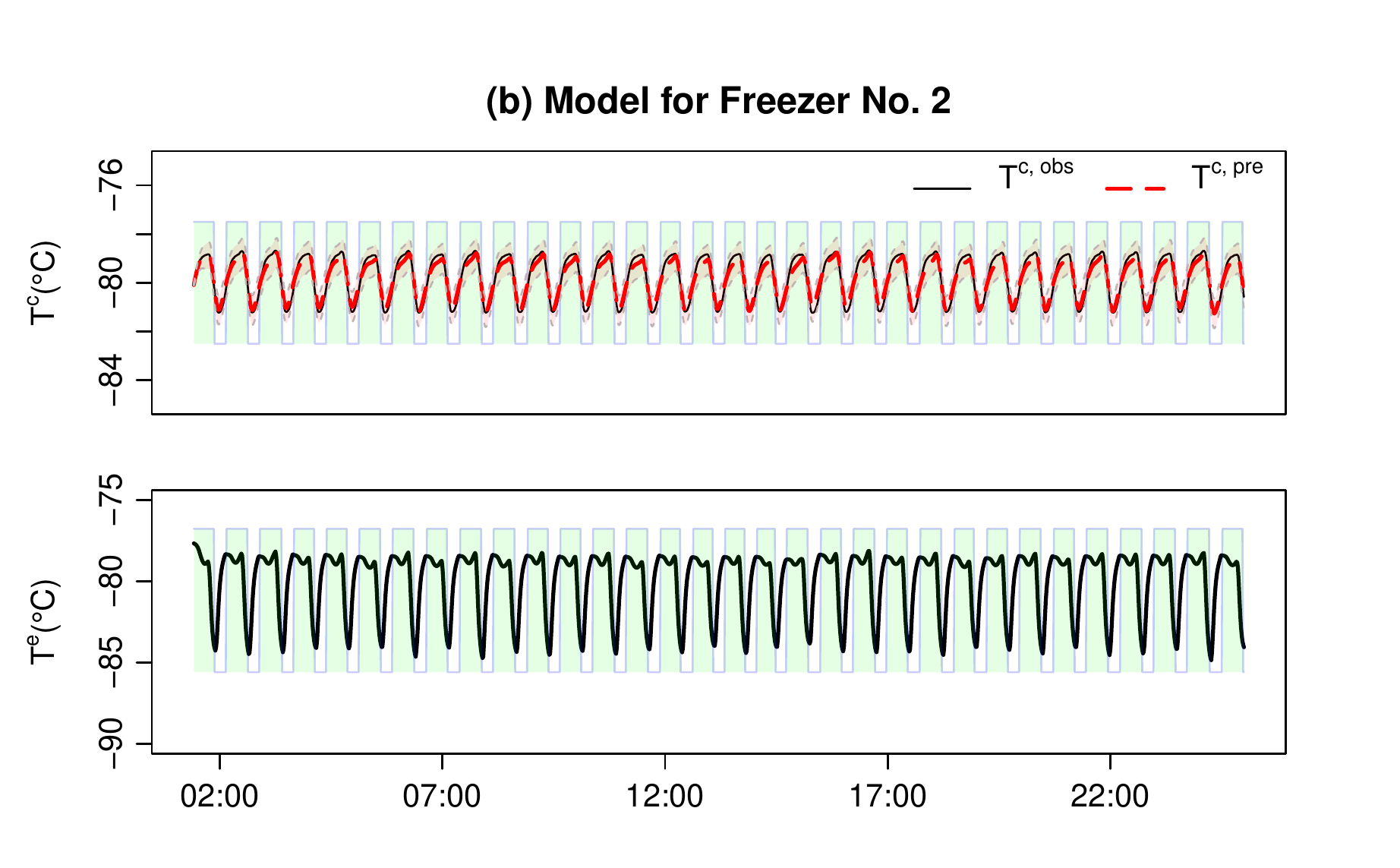}
\caption{Comparisons between the observations ($T^\text{c, obs}$) and the unconditional predictions ($T^\text{c, pre}$) for (a) M$_{F\#1}$ and (b) M$_{F\#2}$. The red shading areas stand for the 95\% confidence intervals of the prediction results. The compressors are running during the square-shaded periods.}
\label{sim}
\end{figure}

\hl{Unconditional prediction is conducted to assess the model predictive performance. Unconditional predictions of the state variables $\bm{T}_t|\bm{T}_0$ are generated by inputting the model with the initial values of three states \{$T^c_0, T^w_0, T^e_0$\} and actual measured input variables \{$T^{a}_t, T^{e, in}_t, T^{e, out}_t, m_t$\} using EKF without updating.} Fig. \ref{sim} shows the \del{unconditional predictions of the }\hl{predicted} RTD temperature from the two freezers. The predictions agree well with the measurements, and the narrow prediction intervals indicate that the predictions are relatively accurate. The root mean squared error for M$_{F\#1}$ and M$_{F\#2}$ are 0.11 $^{\circ}$C and 0.19 $^{\circ}$C, respectively. The results demonstrate that the model captures the most critical dynamics of the freezing chambers. However, the small reduction at 07:00 in F\#1 is not accurately reproduced by the predictions\del{, as they do not show a clear drop}. This reveals that \hl{some underlying dynamics are still not accounted for in the model.}\del{This reveals that there are still some underlying dynamics that are not accounted for in the model.} Identifying these dynamics would require more data containing such infrequent phenomena, which may not be worth considering explicitly. 
\begin{figure}[H]
\centering
\includegraphics[width=1\linewidth]{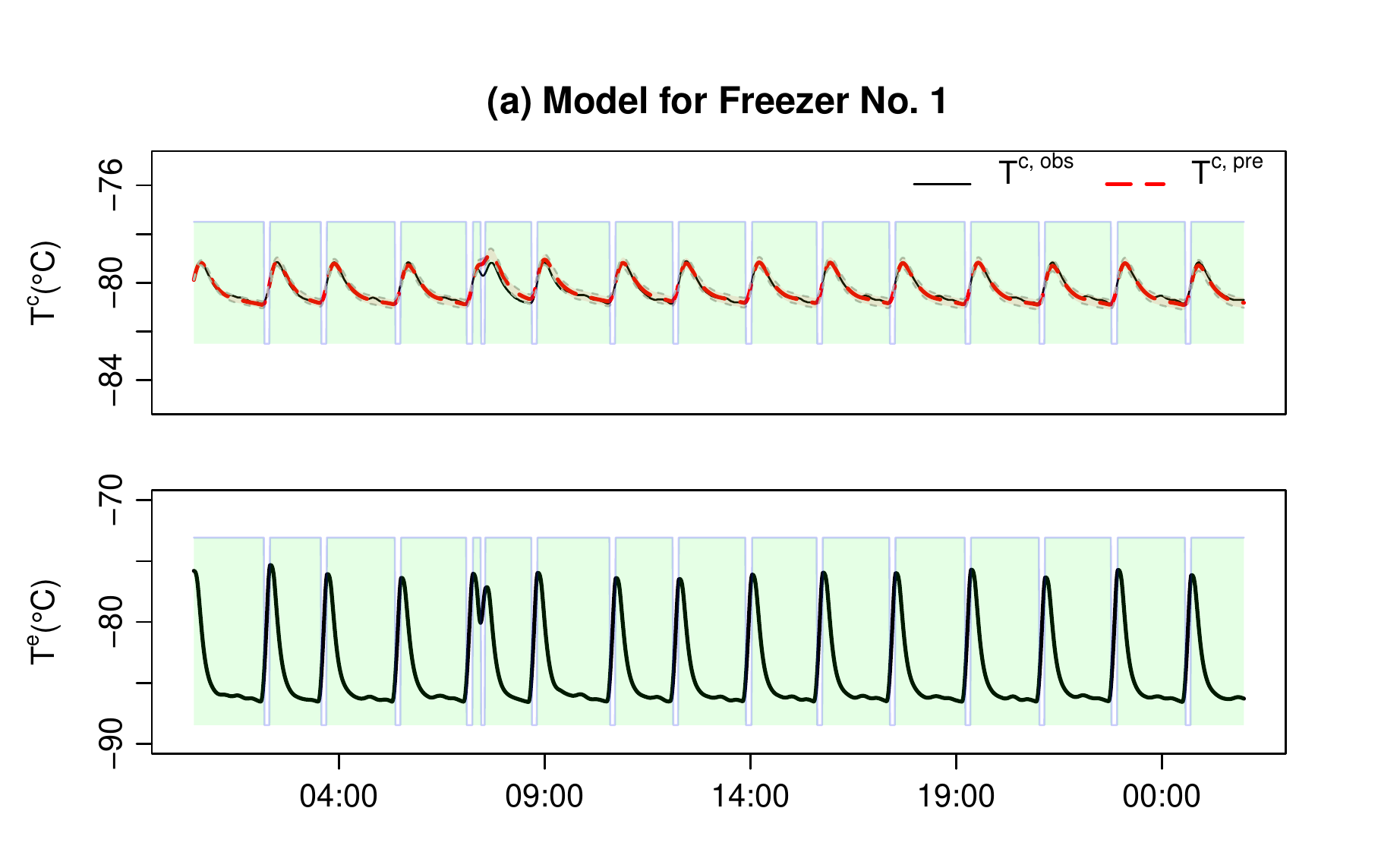}
\end{figure}
\del{Fig. \ref{sim} also presents the predicted local evaporator temperatures, which can provide insight into the underlying heat transfer processes between the evaporator and the freezing chamber.} The predicted local evaporator temperatures are also presented in Fig. \ref{sim} and exhibit distinct patterns for the two freezers. For F\#1, the overall local evaporator temperature evolution closely resembles a classical exponential decay process. It drops as soon as the compressors are ON, indicating that the evaporator is cooled down efficiently. This consequently leads to a fast decay in the RTD temperature. \del{The local evaporator temperature then stabilises at a certain level.}In contrast, the profile of the local evaporator temperature for F\#2 \del{shows a} is complex\del{ pattern}. When the compressors are turned on, the local evaporator temperature undergoes a short-term and slow reduction followed by a slight increase. During this \del{process}\hl{period}, the local evaporator temperature remains relatively stable. Then, a sudden and sharp reduction occurs.\del{, leading to a decrease in the RTD temperature. The duration of the dropping part of the local evaporator temperature is much shorter than the running time of the compressors.} \hl{This pattern may appear to contradict the practical understanding of refrigeration processes but well elucidates the mismatch between the RTD temperature drop and the state signal for F\#2. One possible explanation for this phenomenon is the presence of unmodelled dynamics between the evaporator and the RTD. Hence the model attempts to capture the holistic dynamics of the local evaporator and the unrecognised medium. The unmodeled dynamic process can be, for instance, due to an inadequate synergy between the compressors at the two stages. Another potential reason could be attributed to the thermal couples used to measure the evaporator temperatures. It is possible that either the installation positions are inappropriate or the time constant of the thermal couples is too high to respond adequately to temperature changes. Despite this, the patterns of the RTD temperature indeed demonstrate an inefficient and delayed cooling period during the initial ON states. The modelled local evaporator temperature effectively introduces this delay and limited cooling capacity. In this light, the simulated local evaporator temperature still provides valuable insights into the patterns of the RTD temperature and sheds light on the possible underlying nonlinear heat transfer processes between the refrigeration system and the freezing chamber.}\del{and, to an extent, confirms that the heat transfer between the local evaporator and the chamber is a nonlinear process. However, the reason for such an operational pattern on F\#2 remains unclear.}

\del{The different responses of the local evaporator to the compressor state directly lead to the discrepancy in the responses of the RTD temperature in the modelled ULT freezers. }In previous studies on modelling refrigerators \cite{MASTRULLO201438, ONEILL2014819, SOSSAN20161}, the responses of the chamber temperature to the compressor state signal are generally consistent with what is observed for F\#1. Therefore, the direct use of the state signal in conjunction with global cooling capacity for model identification can be informative. However, the operational pattern for F\#2 is rarely reported and can be challenging to model using a time-invariant system without considering the nonlinear heat transfer. The results demonstrate that the proposed \hl{grey-box} modelling approach can effectively\del{capture such a cryptic phenomenon with limited information. } \hl{capture the most crucial dynamic behaviours of the ULT freezing chambers without requiring detailed knowledge of all the underlying phenomena. Moreover, the prediction results across different operational patterns indicate that the model performance is not ULT freezer dependent, confirming the desired transferability of the proposed modelling approach for various ULT freezers.} \del{ However, performing forecasting requires future inputs, and inaccurate inputs can lead to unreliable forecasts. In the case of the proposed models, obtaining accurate future evaporator inlet and outlet temperatures presents a significant challenge, and this will be the focus of future research efforts.}

\subsection{Residual analysis}\label{sec:acf}
\begin{figure}[H]
\centering
\includegraphics[width=1\linewidth]{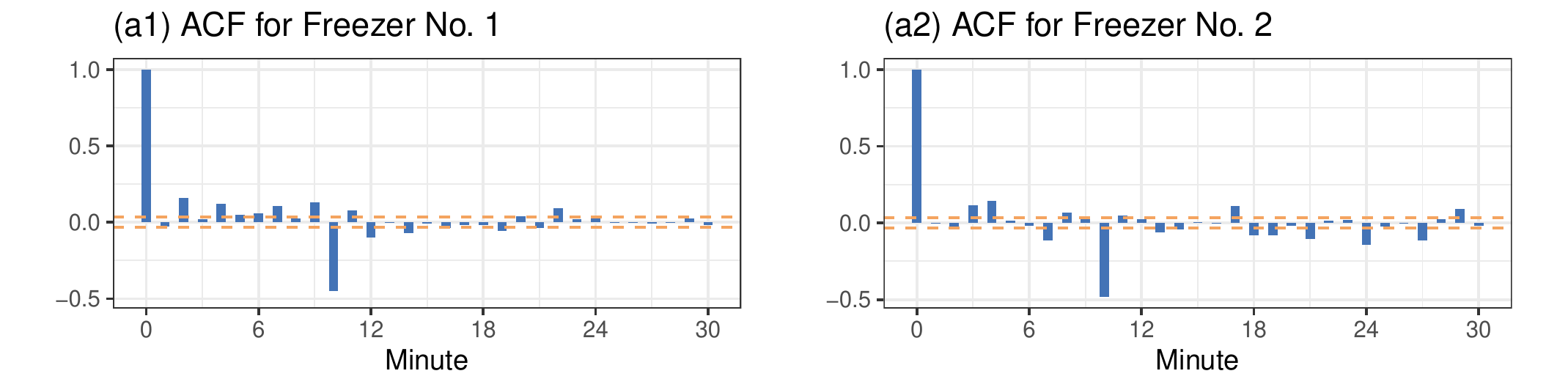}
\includegraphics[width=1\linewidth]{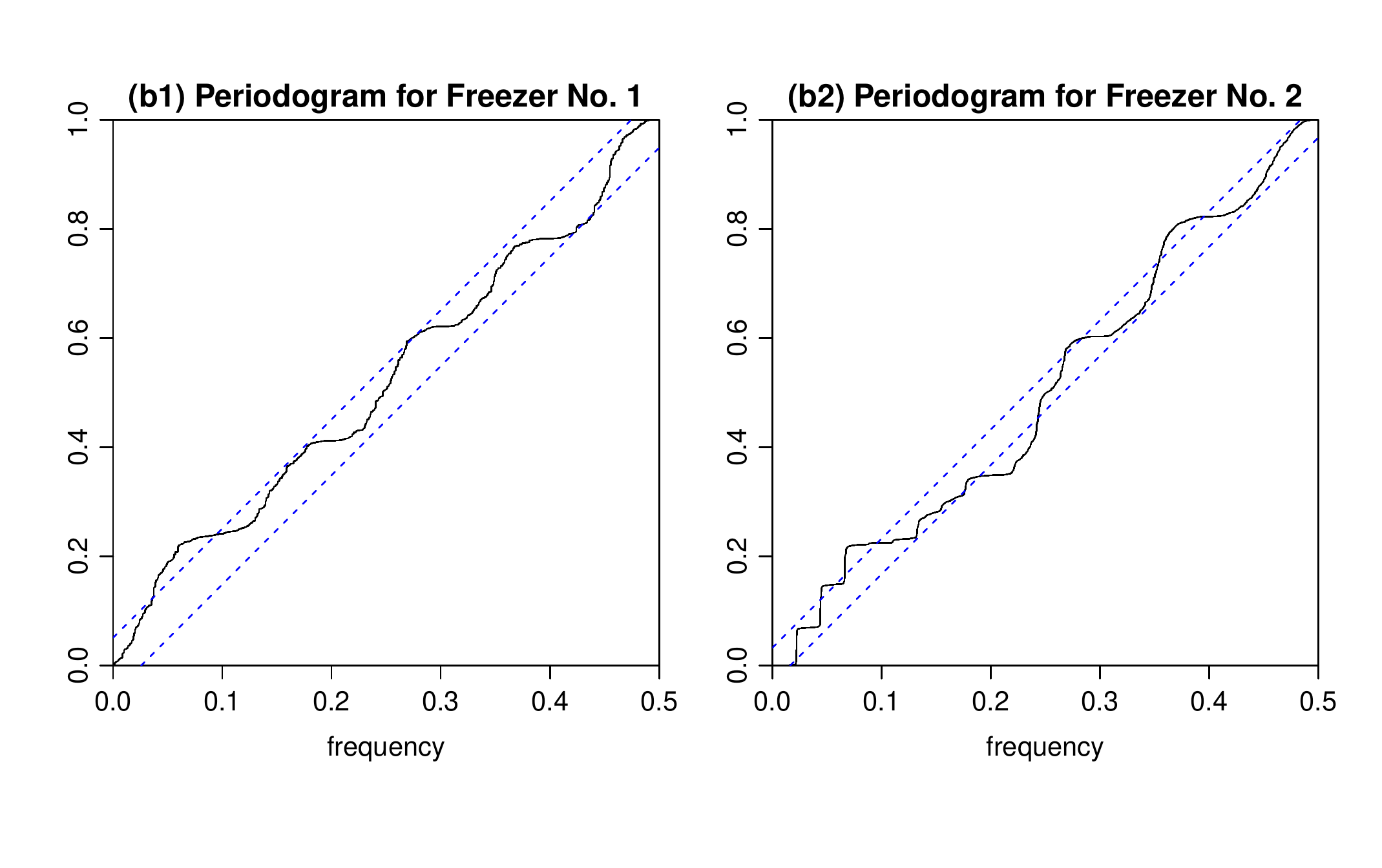}
\caption{(a) The estimated auto-correlation function of the one-step prediction errors and (b) cumulated periodogram for M$_{F\#1}$ and M$_{F\#1}$. The regions between the dashed lines are the 95\% confidence intervals under the hypothesis that the residuals are white noise.}
\label{acf}
\end{figure}

For an SDE-based model, the residuals from the estimation should be uncorrelated to prove that the model describes all auto-correlation of the systems \cite{BACHER20111511}. \del{Fig. \ref{acf} shows the auto-correlation function (ACF) and the cumulated periodograms (CP) of the one-step prediction residuals for M$_{F\#1}$ and M$_{F\#2}$. The residuals from both models are close to being white noise, despite minor spikes in some lags. The CP mostly lie inside of the confidence interval but also shows slight periodicity, implying that some dependency from the inputs still exists.}\hl{As shown in Fig. \mbox{\ref{acf}} (a), the auto-correlation function (ACF) of the one-step prediction residuals for M$_{F\#1}$ and M$_{F\#2}$ are close to being white noise, despite minor spikes in some lags. The cumulated periodograms (CP) presented in Fig. \mbox{\ref{acf}} (b) mostly lie inside of the confidence interval but exhibit slight periodicity, implying that some dependency from the inputs still exists.} Moreover, a strong auto-correlation at lag 10 can be observed in the ACF for both models\del{. This indicates that the way of introducing some of the inputs is not effective in capturing this periodic phenomenon. }\hl{, suggesting that the way of introducing certain inputs does not effectively capture this periodic phenomenon.}\del{The exact reason for this should be supported with further investigation.} \hl{In general, the residual analysis suggests that the models can be relied upon for forecasts, which is the backbone for reliable MPC applications.}
\subsection{Effects of the sigmoid function}\label{sec:sigm}
\begin{figure}[H]
\centering
\includegraphics[width=1\linewidth]{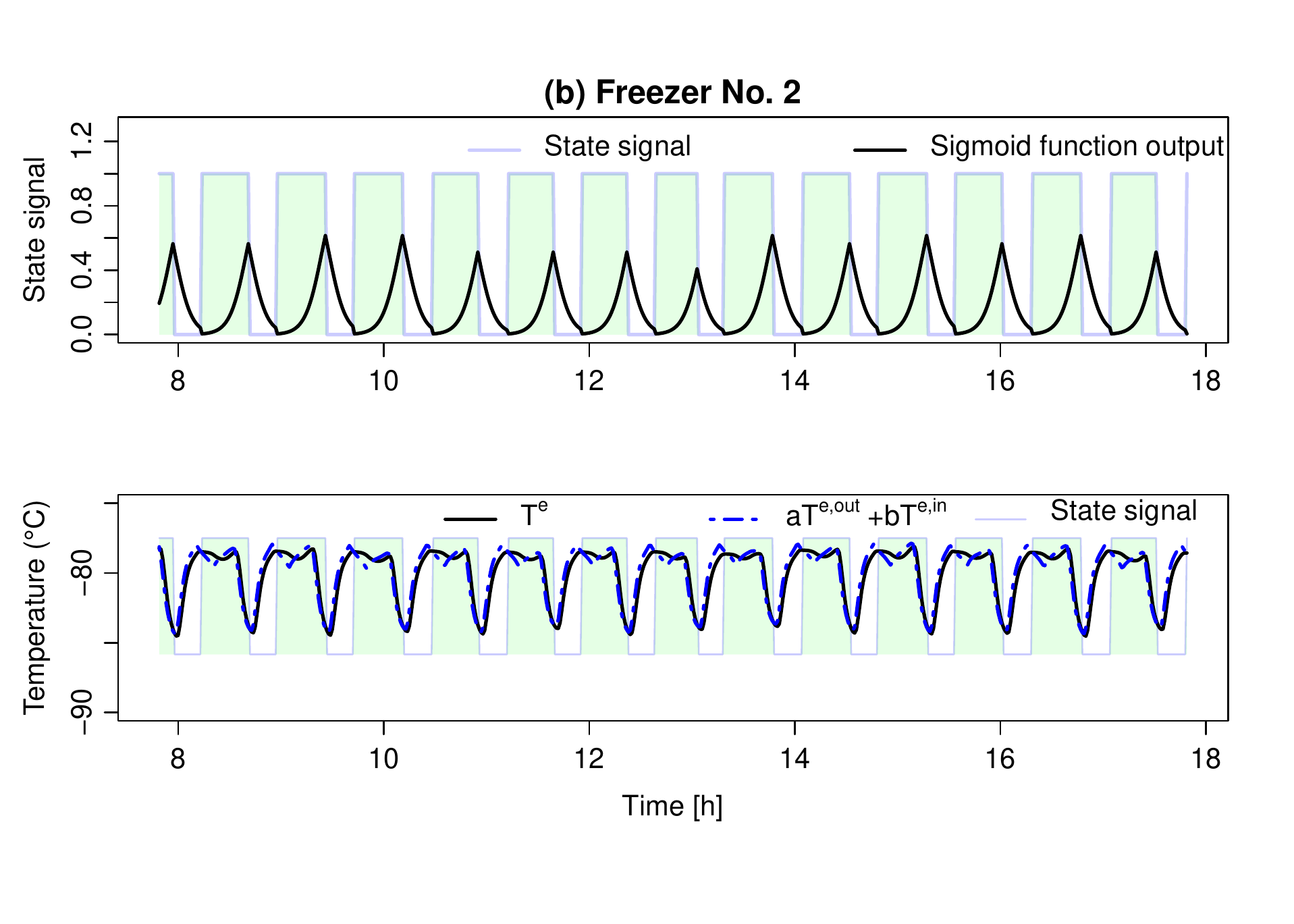}
\caption{Top: Correlation between the compressor state signal and output from the sigmoid function. Bottom: Comparison between the\del{presumed} \mbox{\hl{hypothetical}} evaporator temperature and estimated cold surface temperature. (a) for F\#1 and (b) for F\#2. The compressors are running during the square-shaded periods.}
\label{Te}
\end{figure}

One of the major novelties of this study is to model the underlying nonlinear phenomena in the freezing chamber using a sigmoid function. \del{The sigmoid function modifies the compressor state signal and regulates the changing rate of the local evaporator temperature relative to the\del{presumed} \hl{hypothetical} evaporator temperature.}Fig. \ref{Te} shows the trajectory of the sigmoid function along with the comparisons between the \hl{hypothetical}\del{local} evaporator temperature $aT^\text{e,out}_t$ + $bT^\text{e,in}_t$ and the estimated local evaporator temperature $T^\text{e}$. For F\#1, see Fig. \ref{Te} (a), the sigmoid function transforms the initial part of the ON state to be a smooth increase. During this transition, the increasing rate of the modelled local evaporator temperature is restricted. Due to the relatively short OFF duration, the sigmoid function keeps saturated when the state signal is 0. However, the sigmoid function never reaches 0, confirming that the cooling can be fast transported to the RTD proximity. The bottom plot shows a considerable discrepancy between the \hl{hypothetical} and local evaporator temperatures during the OFF and transition periods. The local evaporator temperature decays more slowly than the \hl{hypothetical one}\del{ evaporator temperature} before the saturated period and becomes nearly identical when the sigmoid function reaches saturation. In addition, the \hl{hypothetical} evaporator temperature peaks between consecutive ON periods. The parameter $C^\text{e}$ buffers this phenomenon such that the peaks of the local evaporator temperature are aligned to the compressor state shifts.
\begin{figure}[H]
\centering
\includegraphics[width=1\linewidth]{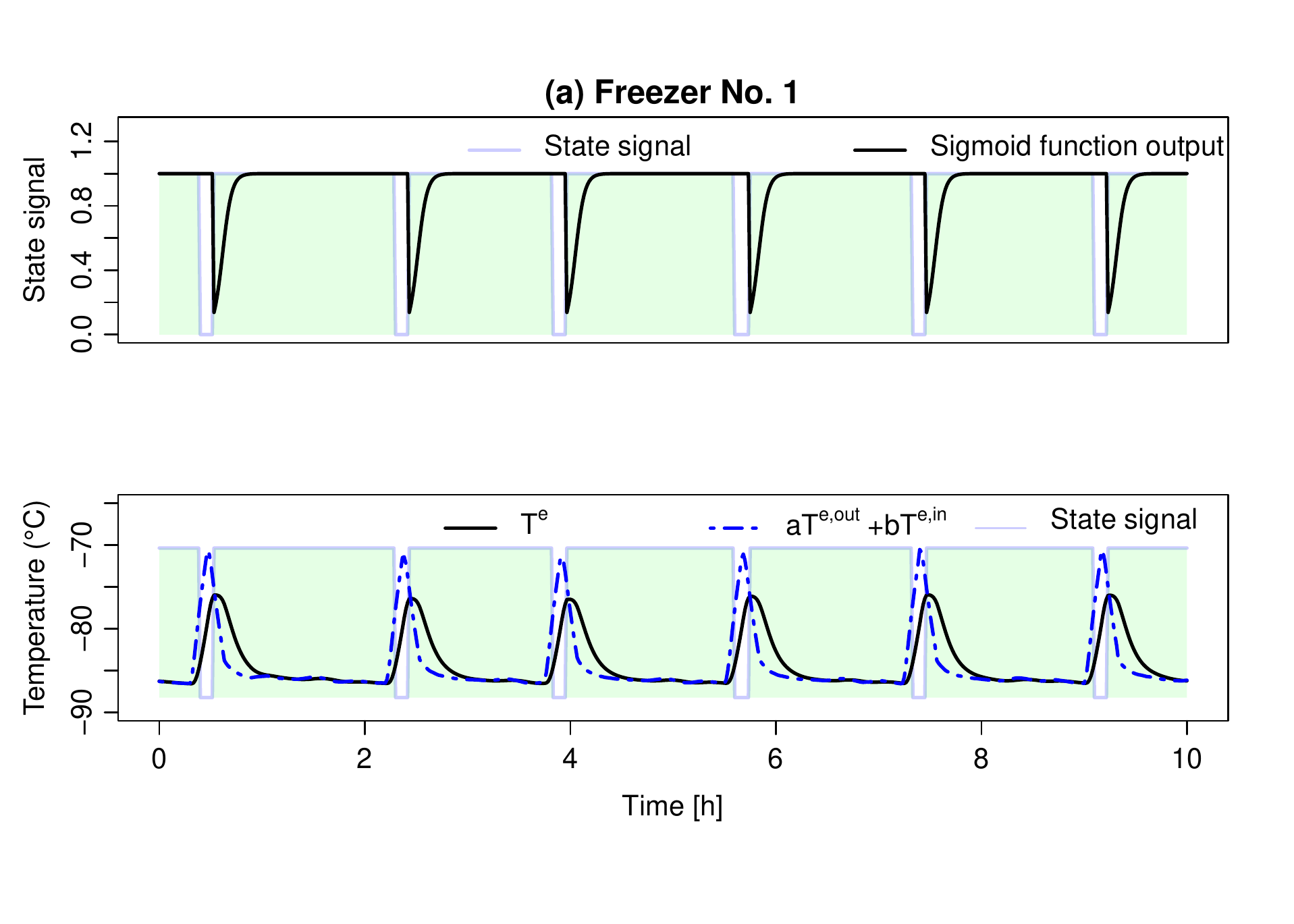}
\end{figure}
F\#2 shows contrasting results compared F\#1, as shown in Fig. \ref{Te} (b). The sigmoid function never reaches saturation, indicating that it always limits the changing of the local evaporator temperature towards the \hl{hypothetical} evaporator temperature. This is confirmed in the bottom plot, in which the local evaporator temperature always slightly deviates from the \hl{hypothetical} evaporator temperature. Although $C^\text{e}$ is small for F\#2, the lagging effect and their discrepancy are still noticeable, especially during the \del{ending }parts \del{of the OFF periods and the initial part of the ON periods }when the output of the sigmoid function is small. In both models, the \hl{hypothetical} and the local evaporator temperature have a similar trend for both freezers, indicating that the linear assumption of the \hl{hypothetical} evaporator temperature is reasonable. \hl{However, it is important to note that the comparison between the hypothetical and local evaporator temperatures is primarily intended to illustrate how the sigmoid function enables efficient prediction of time delays and reduced cooling capacity, while also providing insight into potential underlying phenomena that can be challenging to measure directly. It is not meant to imply that the modelled evaporator temperature represents the ground truth.}

The effects of the sigmoid function are found to play a rather salient role in yielding good model predictions, particularly in the initial stage of the ON period. This confirms that modelling the cooling input as a smooth change \del{rather than a simple step change }is \del{more }appropriate. This can be superficially explained by the fact that thermo-mechanical systems require a warming-up period to reach nominal conditions. Whereas, such warming-up periods may not only depend upon the intrinsic properties of the systems but also be inevitably influenced by other factors such as loading and wear conditions. Therefore, the governing parameters ($\alpha$ and $\beta$) for the sigmoid function may vary for different freezers and even for the same freezers at different lifetimes. This\del{, however, suggests that the model for a specific ULT freezer should be retuned regularly to account for the changes in the system state. It also} reveals that any exact models for ULT freezers can risk limited applicability. In this light, the proposed grey-box modelling approach is scalable and more promising for practical applications. 
\subsection{Model identifiability}\label{sec:ident}
\begin{figure}[H]
\centering
\includegraphics[width=1\linewidth]{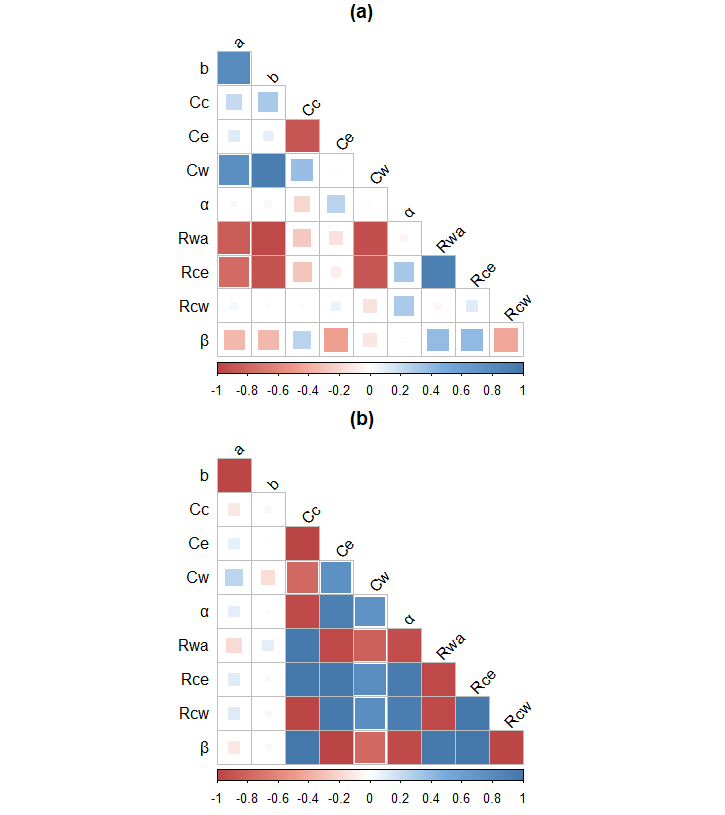}
\caption{Correlogram between estimated parameters. (a) F\#1. (b) F\#2.}
\label{corr}
\end{figure}

\begin{figure}[H]
\centering
\includegraphics[width=0.6\linewidth]{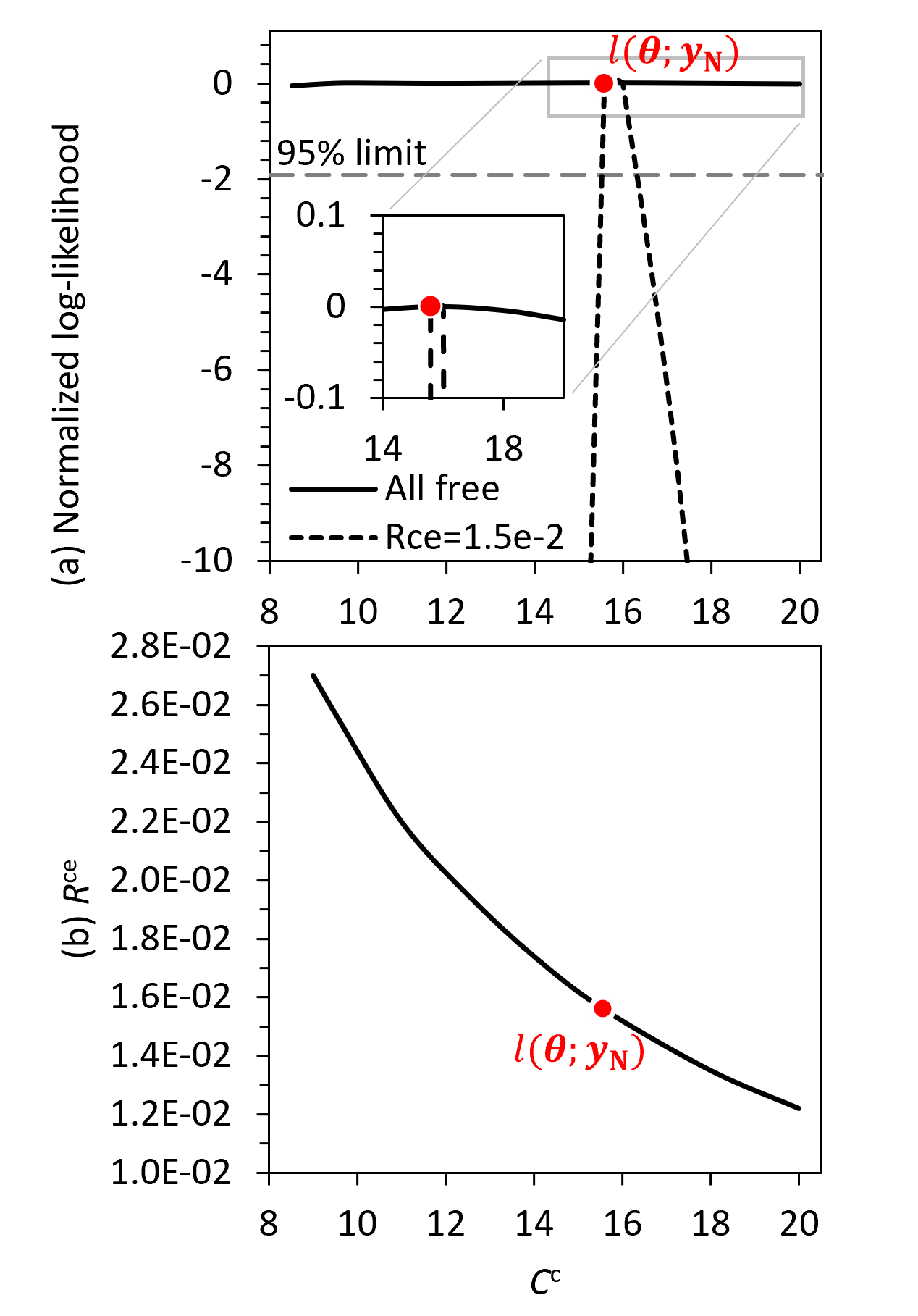}
\caption{(a) Profile likelihood estimated at different $C^\text{c}$ with all other parameters  remaining free and with $R^\text{ce}=1.5e-2$. (b) Correlation between $C^\text{c}$ and $R^\text{ce}$ derived at the maximum log-likelihood.} 
\label{pl}
\end{figure}

Fig. \ref{corr} displays the correlogram of the estimated parameters. The patterns of the correlation matrices for the two models are different, but fairly strong correlations ($|R| > 0.9$) can be found between some parameters. This is particularly clear for M$_{F\#2}$, in which strong correlations can be observed between thermal-related parameters $C$ and $R$. It indicates that these parameters are not independent, and the model may not be practically identifiable.  
We employ the profile likelihood to \hl{further} evaluate the identifiability. Because of computationally burdensome, we \del{use M$_{F\#2}$ as a demonstration and }only select $C^\text{c}$ \hl{from M$_{F\#2}$} as the parameter of interest because it is highly correlated with other parameters. Fig. \ref{pl} shows the normalized profile log-likelihood for $C^\text{c}$ in a solid line. The profile likelihood distribution is rather flat and always above the 95\% confidence threshold, despite a reducing tendency at large $C^\text{c}$. This indicates that a similar level of maximum likelihood can be derived from a wide range of $C^\text{c}$. The dashed line in Fig. \ref{pl} plots the profile likelihood of $C^\text{c}$ by keeping $R^\text{ce}$ to be the estimated value 1.5e-2. The profile likelihood distribution shows a well-defined confidence interval. This suggests that $C^\text{c}$ and $R^\text{ce}$ are dependent. This mutual dependency is confirmed by Fig. \ref{pl} \hl{(b)}, in which a clear relationship between the $C^\text{c}$ and $R^\text{ce}$ at the maximum log-likelihood is shown. The results suggest that the model is not practically identifiable and imply that the classic quadratic assumption does not strictly hold for all parameters.
Identifiability is a crucial prerequisite to reliable parameter inference. The weak identifiability of the proposed models is mainly due to the absence of a dimension constraint in energy quantity. \del{This results in a strong parameter correlation. }Additionally, the limited variability in the non-intrusive data used for identification can also contribute to weak identifiability, as shown in Fig. \ref{F1}. The non-intrusive data may contain too little information \hl{on the system dynamics} to achieve the unicity of each parameter estimation. Therefore, we should avoid interpreting them as the exact physical proprieties of the systems. 

The results in Fig. \ref{pl} (b) point out a possible way to enhance the identifiability of the model, that is, to fix some of the parameters to be the true value to reduce the degree of the freedoms. This, however, requires domain knowledge and may entail demand for extra experiments. \hl{The less strong correlations between the parameters of M$_{F\#2}$ somewhat indicate more reliable parameter estimates, reflecting that specific dynamical patterns can also enhance the model identifiability.} Another approach to improve the level of data information is through excitation of the system against a large spectrum of frequencies using a periodic and deterministic pseudo-random binary sequence \cite{SOSSAN20161, MADSEN199567, BACHER20111511, YU2021110775}. However, this also requires controlled experiments and thus is not always accessible.  

Although a fully interpretable model is desirable, it can be demanding to build and \del{maintain in practice}sustain in practical settings. Despite the lack of identifiability of the present model, it effectively captures the main heat dynamics of the system and can easily adapt to the system state changes through\del{streaming-data-based} parameter re-estimation \hl{based on operational data}. This feature strongly facilitates practical applications and maintenance of the model.
\subsection{\del{Relative parameter interpretation}Long-term model performance}\label{sec:pire}
\begin{figure}[H]
\centering
\includegraphics[width=1\linewidth]{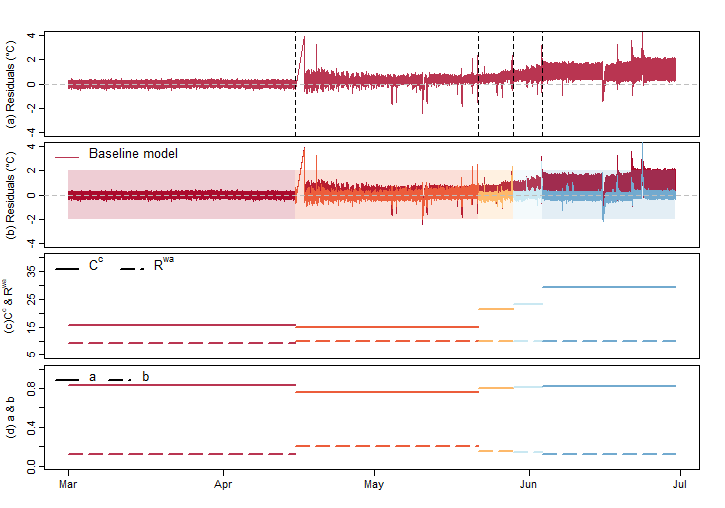}
\caption{(a) Temporal evolution of residuals from the baseline M$_{F\#2}$ from March to July. (b) Comparisons between the residuals from the baseline model and four retuned models. (c-d) retuned parameters for each model.}
\label{infer}
\end{figure}

\hl{In practical operations, ULT freezers are subject to continuous external disturbances, which can lead to changes in chamber thermal properties and alter the dynamical behaviours. For instance, the heat capacity of the chamber can change due to loading activities. Frosting on the inner surface of the chamber can increase the thermal resistance. Consequently, a specific model will not maintain throughout the entire lifespan of an ULT freezer. It thus becomes necessary to update the model to capture the evolving heat dynamics of the freezing chamber when the previous dynamical equilibrium is disrupted.

Long-term unconditional predictions are conducted to elaborate on this issue. We selected a 4-month regular operational period of F\#2, which included multiple door-opening events. Fig. \mbox{\ref{infer}} (a) shows the unconditional prediction residuals from the baseline model M$_{F\#2}$ over the 4 months. The residuals are centred along zero within a narrow span during the first 1.5 months when no external disturbance occurs. Afterwards, the residuals start to drift from zero, suggesting that the baseline model M$_{F\#2}$ underestimates the RTD temperature. Therefore, relevant parameters should be retuned to maintain the model performance. }
We \hl{then} selected four events and divided the 4-month into five sub-periods, separated by the vertical dashed lines, see Fig. \ref{infer} (a). Apart from the first reboot event, all events were door-opening. We chose them\del{ four events} because the residuals exhibit a clear tendency to drift from the previous level\del{ after each event. This indicates that the dynamic equilibrium of the freezers has changed, and the parameter should be retuned}. There are also a few door-opening events within each sub-period reflected by the spikes in the residuals. However,\del{ as discussed earlier,} they do not cause significant drift on the residuals and are thus not considered to demand parameter updates.

As discussed in Subsection \mbox{\ref{sec:pe}}, the $C^\text{w}$, $C^\text{e}$, $R^\text{ce}$, and $R^\text{wa}$ values \hl{of M$_{F\#2}$} are more likely to be related to the intrinsic properties of the freezers, while $C^\text{c}$ and $R^\text{cw}$ are more easily affected by external disturbances. \hl{Thus, it is reasonable to retune only $C^\text{c}$ and $R^\text{cw}$ while fixing the remaining ones.} $a$ and $b$ values also need to be retuned to adjust the contribution of the inlet and outlet temperatures to the local evaporator temperature. This partial retuning strategy can reduce computational time and improve the model identification stability\del{, thereby making the updated parameters easier to interpret}.
\del{Because the proposed model is built upon the heat transfer theory, the relative changes in the parameters enable the possibility of physically interpreting the parameters relative to their past states, notwithstanding the lack of physical meanings. For example, it has been proven that M$_{F\#1}$ and M$_{F\#2}$ can effectively predict the RTD temperature for F\#1 and F\#2. In the absence of external disturbances, both models are expected to accurately track the freezers' state. The model can be used as a baseline model. However, when events occur, say door-opening with loading activities, the parameters related to the thermal properties may change due to a shift in dynamic equilibrium. Thus the relevant parameters should be retuned to maintain the performance. The model with the retuned parameters will serve as the new baseline model. For a typical-sized ULT freezer, changes in content mass in the chamber are not expected to be significant at each loading event. This means the previous baseline model may only need to be retuned after several loading events. Furthermore, it is reasonable to retune only a few parameters while fixing the remaining ones. Taking M$_{F\#2}$ as an example, the $C^\text{w}$, $C^\text{e}$, $R^\text{ce}$, and $R^\text{wa}$ values are more likely to be related to the intrinsic properties of the freezers, while $C^\text{c}$ and $R^\text{cw}$ are more easily affected by external disturbances. Likewise, $a$ and $b$ values also need to be retuned to adjust the contribution of the inlet and outlet temperature to the local evaporator temperature. This strategy reduces computational costs and \hl{improves the model identifiability, thereby making}  the updated parameters easier to be interpreted.}

\del{we conducted a demonstrative study to elaborate on this strategy.  We thus selected a 4-month regular operational period from F\#2 that included door-opening events in order to illustrate how the critical parameter can possibly be interpreted to provide a relative implication for the operation status. Fig. \ref{infer} (a) shows the unconditional prediction residuals from the baseline model M$_{F\#2}$ during the 4 months. The residuals are well-centred along zero within a narrow span at the beginning of 1.5 months, during which no external disturbance occurs. Afterwards, the residuals start to drift from zero, and several spikes can be observed until the end of the time series. The baseline model will continuously underestimate the RTD temperature if it is not retuned. The consistently increasing trend of the residuals somewhat implies that the characteristics of the door-opening events are the same, i.e., either loading or unloading.}

Fig. \ref{infer} (b) presents the unconditional prediction residuals from the four updated models during each sub-period. \del{Compared to the baseline model, t}The retuned models lead to a similar level of residuals \hl{as the baseline model}\del{ centred around zero}. However, the residuals still spike at \del{certain points during }door-opening events. This is expected as the model does not account for infiltration and thus cannot effectively predict temperature evolution during door-opening events\del{ and the consecutive pull-down period. However, modelling infiltration in a simplistic manner can be challenging because it }\hl{, which can heavily depend on random }user behaviours \del{, such as opening duration and opening ratio }\cite{MASTRULLO201438}. Moreover, it may be less critical to reproduce the temperature profile during door-opening events, especially if these events are being logged.

Fig. \ref{infer} (c) shows the retuned $C^\text{c}$ and $R^\text{wa}$ values. $C^\text{c}$ is relatively more sensitive to state changes.\del{ The increasing levels of $C^\text{c}$ \hl{possibly} indicate that the mass of the contents in the chamber increases after each selected event, apart from the first update, where a slightly lower $C^\text{c}$ is derived. By having access to detailed information about the properties of the loaded contents, it could be possible to quantify such mass increases based on the percentage changes in $C^\text{c}$. As discussed, the absolute values of $C^\text{c}$ may not have direct physical meaning, but the relative changes can be interpreted as the total mass in the chamber increasing by -4\%, 44\%, 8\%, 26\% from its previous level, assuming the specific heat capacity of the loaded contents is similar.} In contrast, the level of $R^\text{cw}$ changes insignificantly. Fig. \ref{infer} (d) shows the retuned $a$ and $b$ values. The results suggest again that $a$ and $b$ are negatively correlated.\del{Despite an unchanged $C^\text{c}$ value, t} The estimated $a$ is smaller after the first event and gradually increases thereafter\del{to a similar level as the undisturbed period}.\del{ The smaller $a$ for the 1$^\text{st}$ retuned model could be attributed to  the different content spatial distribution in the chamber.} Overall, the changes in $a$ and $b$ are small.\del{, which is expected in practice.}

\del{Fig. \ref{infer} (c) shows the retuned $C^\text{c}$ and $R^\text{wa}$ values. $C^\text{c}$ is relatively more sensitive to state changes.\del{As discussed, the absolute values of $C^\text{c}$ may not have direct physical meaning, but the relative changes can be interpreted as the total mass in the chamber increasing by -4\%, 44\%, 8\%, 26\% from its previous level, assuming the specific heat capacity of the loaded contents is similar.} In contrast, the level of $R^\text{cw}$ changes insignificantly.

Fig. \ref{infer} (d) shows the retuned $a$ and $b$ values. The results suggest again that $a$ and $b$ are negatively correlated. \del{Despite an unchanged $C^\text{c}$ value, t}The estimated $a$ is smaller after the first event and gradually increases thereafter\del{to a similar level as the undisturbed period}. The smaller $a$ for the 1$^\text{st}$ retuned model could be attributed to \del{ the different content spatial distribution in the chamber. }\hl{a reduced contents in the chamber or specifically in the upper part of the chamber.} Overall, the changes in $a$ and $b$ are small, which is expected in practice.}

\hl{The results demonstrate that the freezers are subject to unpredictable changes in their dynamic behaviours over time. Therefore, updating the model is necessary to sustain its performance in the long run. This highlights the advantage of using a stochastic grey-box model, which allows the parameters to be easily tunned using measured time series data while considering stochasticity. Such flexibility greatly facilitates resilient online applications and decision-making under uncertainty.}

\hl{Furthermore, the parameter retuning processes may provide an alternative approach to physical interpretation. The proposed model is built upon the heat transfer theory, ensuring the validity of the simulated heat transfer processes and the roles of the parameters within the model structure. Consequently, the parameters should function according to their definitions. Even though the absolute values of the parameters may lack meaning, their relative changes can potentially be interpreted based on their physical definitions. For instance, increasing levels of $C^\text{c}$ in Fig. \mbox{\ref{infer}} (c) possibly indicate an increased mass of the contents in the chamber after each selected event. Changes in $a$ and $b$ can to an extent reflect the evaporation status in the evaporator. This relative parameter inference strategy can also offer valuable information on the operational status of ULT freezers and is promising to compensate for the low model identifiability. Unfortunately, the lack of detailed information about loading events hinders the validation of this strategy against the actual events. Experiments are necessary to confirm the effectiveness of this approach.}
\del{using the strategy that interprets the parameter relative to the previous state. This strategy leads to a semi-interpretable model, which helps to compensate for the lack of exact physical meaning in the parameters. The trade-off between modelling complexity and interpretability is thus addressed. The relative changes in the parameters can provide useful information for operators to remotely monitor the health status of the ULT freezer. For instance, when the demand for parameter update arises without any registered man-made event, it may indicate the presence of faults. A similar conclusion can be drawn when a retuned parameter shows abnormal changes. Therefore, the model has the potential to function as a fault detection system for ULT freezers. Integrating a fault diagnostics algorithm will allow for the full functionality of FDD.}
\subsection{Potential applications and limitations}
\hl{The developed modelling approach demonstrates strong prediction performance and transferability, serving several potential digital applications. One such application is continuous online surveillance, proven by model predictive performance and convenient parameter re-estimates. Consequently, the model can function as a residual generator through Kalman Filter for FDD purposes. Studies have reported that the innovation from the Kalman Filter can effectively detect subtle changes in the system states \mbox{\cite{YANG201113233, OKATAN2009762, Jieyang}}, which are challenging to identify through simple setpoint alarms alone. This capability contributes to early fault detection, leading to timely fault correction. Furthermore, good predictive performance and residual analysis evidence the potential for forecasting tasks. This ability makes the model qualified for MPC purposes, which is promising to reduce energy consumption from ULT freezers.}

Several limitations of the present study must be made clear. First, the data used for the modelling processes comes from the ULT freezers with 2CRS. It is unclear whether the proposed modelling approach can be extrapolated to ULT freezers with other refrigeration systems, e.g. Auto-Cascading Refrigeration Systems \cite{review}. Generalizing the results to other types of ULT \hl{and standard freezers} would\del{ be interesting and could }help to improve the understanding of their dynamic behaviour. Second, the availability of the evaporator inlet and outlet temperatures may vary among different brands of ULT freezers. However, it is increasingly common for multiple sensors to be installed in ULT freezers by several manufacturers. Also, measuring temperatures remain relatively easy and inexpensive \hl{that can be feasibly implemented in practice to facilitate the model applications.} Thirdly, \del{due to the lack of detailed information about loading events, it is impossible to validate the results and conclusions of Subsection \ref{sec:pire} against the actual events.}\hl{the relative parameter interpretation strategy suggested in Subsection \mbox{\ref{sec:pire}} is not validated.} A case study \del{investigating the ability of real-time monitoring and fault detection using relative parameter interpretation would be worth conducting}\hl{ would be worthwhile to evaluate its effectiveness and practical applicability.}
\section{Conclusions}\label{sec:conclusions}
A reliable dynamic model is essential for intelligent surveillance and \del{smart}\hl{efficient} energy management for ULT freezers. This study develops a novel approach to modelling the heat dynamics of the ULT freezing chamber based exclusively on costless data from the embedded sensors. The model consists of two fully linear states, RTD temperature and chamber wall temperature. In addition, a hidden state of the local evaporator temperature is formulated based on a sigmoid function and a modified compressor state signal, enabling for effective incorporation of time delays and reduced cooling capacity into the system. The established models are proven to \del{effectively}\hl{accurately} capture the dynamics of the RTD temperature in ULT freezers with different operational patterns. The unconditional predictions are in good agreement with the measurements, \del{suggesting that the model can be 
used for continuous monitoring of the chamber temperature and is highly promising for fault detection and MPC implementations.}\hl{demonstrating its potential for continuous chamber temperature monitoring.}

The proposed modelling approach significantly promotes the practical applicability and transferability of grey-box modelling of ULT freezing chambers, as no costly meters and intrusive experiments are required. \hl{More importantly, the ability to easily retune the parameters while considering uncertainties over time enables the model to maintain its reliable prediction performance throughout the entire lifespan of the ULT freezers.}\del{However, it has a major drawback of lacking identifiability, making it difficult to associate the model parameters with any physical meaning.} \hl{However, a major limitation is the low model identifiability, which hampers the direct association of model parameters with physical meaning.} \hl{ Addressing this issue by analyzing the relative changes observed during each model update holds promise and warrants further investigation. Future research will also focus on practical implementations of the model, such as FDD and MPC.} \del{Despite this, critical parameters can potentially be interpreted relative to their past levels through continuous parameter updates. The relative changes in the retuned parameters can also}
\section*{CRediT authorship contribution statement}
\textbf{Tao Huang}: Conceptualization, Methodology, Software, Validation, Investigation, Visualization, Writing - Original Draft. \textbf{Peder Bacher}: Conceptualization, Supervision, Writing - review \& editing, Funding acquisition, Project administration. \textbf{Jan Kloppenborg Møller}: Conceptualization, Supervision, Writing - review \& editing. \textbf{Francesco D'Ettorre}: Resources, Writing - review \& editing. \hl{\textbf{Wiebke Brix Markussen}: Resources, Writing - review \& editing, Project administration}. 

\section*{Declaration of Competing Interest}
The authors declare that they have no known competing financial interests or personal relationships that could have appeared to influence the work reported in this paper.
\section*{Acknowledgment}
This work was supported by the Danish Energy Technology Development and Demonstration Program (Grant No. 64021-1035). The authors appreciate \del{Wiebke Brix Markussen,}Frederik Wulff Winthereik, and Jakob Thomsen from Danish \hl{Technological} Insititute for their support during the data collection.
\appendix
\section{Calculation of correlation matrix}\label{apped:a}
\hl{Because the Maximum Likelihood estimator is an asymptotically Gaussian with mean $\bm{\theta}$ and covariance}
\begin{equation}
    \bm{\Sigma_{\hat{\theta}}} = \bm{H}^{-1},
\end{equation}
\hl{where, $\bm{H}$ is the Hessian evaluated at the minimum of the objective function, i.e.}
\begin{equation}
\{h_{ij}\} = -(\pdv{}{\theta_i,\theta_j}\mathcal{l}(\bm{\theta};\bm{y}_N))|_{\bm{\theta}= \bm{\hat{\theta}}}, i,j = 1, ..., p
\end{equation}
\hl{The correlation matrix $\bm{R}$ can be obtained by decomposing the covariance matrix} 
\begin{equation}
     \bm{\Sigma_{\hat{\theta}}} = \bm{\sigma_{\bm{\hat{\theta}}}}\bm{R}\bm{\sigma_{\bm{\hat{\theta}}}},
\end{equation}
\hl{where, $\bm{\sigma_{\bm{\hat{\theta}}}}$ is the diagonal matrix of the standard deviations of the estimated parameters.}

\bibliographystyle{elsarticle-num-names} 
\bibliography{elsarticle-template-num}





\end{document}